\documentclass[sigconf]{acmart}

\newcommand{\ignore}[1]{}
\usepackage{enumitem}
\usepackage{multirow}
\usepackage{physics}
\usepackage{amsmath}
\usepackage[noend,linesnumbered,ruled,vlined]{algorithm2e}
\usepackage{tikz}
\newcommand*\circled[1]{\tikz[baseline=(char.base)]{
            \node[shape=circle,draw,inner sep=0.5pt] (char) {#1};}}

\def\BibTeX{{\rm B\kern-.05em{\sc i\kern-.025em b}\kern-.08em
    T\kern-.1667em\lower.7ex\hbox{E}\kern-.125emX}}
\definecolor{aliceblue}{rgb}{0.94, 0.97, 1.0}
\usepackage{xcolor}
\usepackage{tcolorbox}
\tcbset{width=1\columnwidth,boxrule=1pt,colback=aliceblue,arc=1pt,left=2pt,right=2pt,boxsep=0.5pt}

\AtBeginDocument{%
  \providecommand\BibTeX{{%
    \normalfont B\kern-0.5em{\scshape i\kern-0.25em b}\kern-0.8em\TeX}}}

\setcopyright{acmcopyright}
\copyrightyear{2021} 
\acmYear{2021} 
\setcopyright{acmcopyright}\acmConference[MICRO '21]{MICRO'21: 54th Annual IEEE/ACM International Symposium on Microarchitecture}{October 18--22, 2021}{Virtual Event, Greece}
\acmBooktitle{MICRO'21: 54th Annual IEEE/ACM International Symposium on Microarchitecture (MICRO '21), October 18--22, 2021, Virtual Event, Greece}
\acmPrice{15.00}
\acmDOI{10.1145/3466752.3480059}
\acmISBN{978-1-4503-8557-2/21/10}

\begin{document}

\title{ADAPT: Mitigating Idling Errors in Qubits\\ via Adaptive Dynamical Decoupling}


\author{Poulami Das}
\authornote{Both authors contributed equally to this research. The corresponding authors can be be reached at poulami@gatech.edu and stannu@wisc.edu.}
\affiliation{%
  \institution{Georgia Tech}
  \city{Atlanta}
  \country{USA}
}

\author{Swamit Tannu}
\authornotemark[1]
\affiliation{%
  \institution{University of Wisconsin}
  \city{Madison}
  \country{USA}}

\author{Siddharth Dangwal}
\affiliation{%
  \institution{IIT Delhi}
  \city{New Delhi}
  \country{India}
}

\author{Moinuddin Qureshi}

\affiliation{%
  \institution{Georgia Tech}
  \city{Atlanta}
  \country{USA}
}

\renewcommand{\shortauthors}{Das, Tannu, Dangwal, and Qureshi}

\begin{abstract}
The fidelity of applications on near-term quantum computers is limited by hardware errors. In addition to errors that occur during gate and measurement operations, a qubit is susceptible to {\em idling errors}, which occur when the qubit is idle and not actively undergoing any operations. To mitigate idling errors, prior works in the quantum devices community have proposed {\em Dynamical Decoupling (DD)}, that reduces stray noise on idle qubits by continuously executing a specific sequence of single-qubit operations that effectively behave as an identity gate. Unfortunately, existing DD protocols have been primarily studied for individual qubits and their efficacy at the application-level is not yet fully understood.

Our experiments show that naively enabling DD for every idle qubit does not necessarily improve fidelity. While DD reduces the idling error-rates for some qubits, it increases the overall error-rate for others due to the additional operations of the DD protocol. Furthermore, idling errors are program-specific and the set of qubits that benefit from DD changes with each program. To enable robust use of DD, we propose {\em Adaptive Dynamical Decoupling (ADAPT)}, a software framework that estimates the efficacy of DD for each qubit combination and judiciously applies DD only to the subset of qubits that provide the most benefit. ADAPT 
employs a {\em Decoy Circuit}, which is structurally similar to the original program but with a known solution, to identify the DD sequence that maximizes the fidelity. To avoid the exponential search of all possible DD combinations, ADAPT employs a localized algorithm that has linear complexity in the number of qubits. Our experiments on IBM quantum machines (with 16-27 qubits) show that ADAPT improves the application fidelity by 1.86x on average and up-to 5.73x compared to no DD and by 1.2x compared to DD on all qubits. 
\end{abstract}




\maketitle

\section{Introduction}
\label{sec:intro}
Quantum hardware available today with fifty-plus qubits can already outperform the world's most advanced supercomputer for certain problems~\cite{QCSup}. In the near-term, we can expect quantum computers with few hundreds of qubits to solve certain domain-specific applications~\cite{qaoa1,emani2019quantum,qaoa,vqe,orus2019quantum}. Unfortunately, the fidelity of applications executed on these {\em Noisy Intermediate Scale Quantum (NISQ) }~\cite{preskill2018quantum} computers is limited by the high error-rates of the physical qubit devices. The probability of encountering an error on NISQ computers increases with the size of the program. Therefore, developing software solutions that can reduce the impact of hardware errors and improve the fidelity of NISQ applications is an active area of research.

A qubit can encounter errors while performing gate or measurement operations. Additionally, a qubit can also accumulate errors while it is idle and not performing any operations. These errors, referred to as {\em idling errors}, are observed on both superconducting~\cite{IBMDD,chen2021exponential} and ion-trap hardware~\cite{pokharel2018demonstration}. Under certain circumstances, the idling error-rate of a qubit can exceed the error-rate from gate operations. Furthermore, idling errors can increase significantly in the presence of other active qubits in the vicinity, as the idle qubit accumulates phase noise due to crosstalk generated by the on-going gate operations. Our experiments on IBMQ hardware show that an idle qubit is almost 10x more vulnerable to errors when two-qubit gate operations are scheduled adjacent to it. Thus, idling errors can significantly degrade the fidelity of quantum programs and we focus on mitigating these errors in this paper.

\begin{figure*}[t]
\centering
    \includegraphics[width=\textwidth]{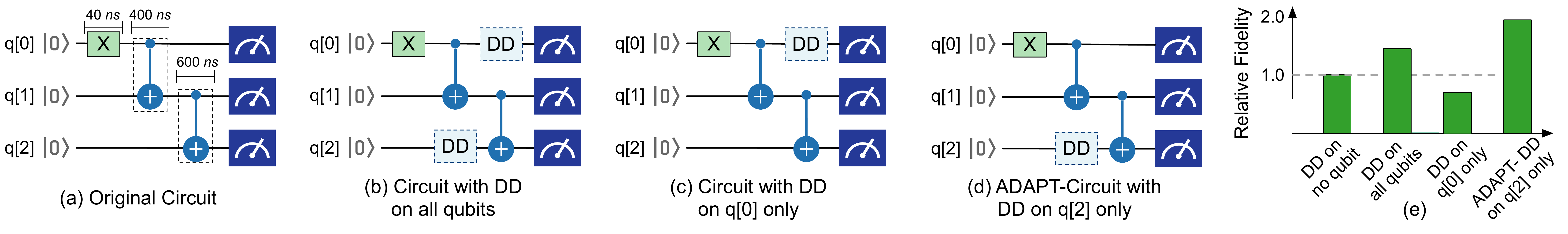}
    \caption{(a) Baseline circuit -- qubits q[0] and q[2] have significant idle time, whereas q[1] remains busy (b) Applying DD on all idle qubits (c) Applying DD on qubit q[0] (d) Applying DD on qubit q[2]  (e) Reliability of DD sequences compared to no DD (Note: Figure is for illustration purposes only). } 
    \label{fig:intro}
    \vspace{-0.05in}
\end{figure*}

The characteristics of quantum programs and NISQ hardware cause a large number of program qubits to remain idle during the execution. There are three key reasons for qubits to remain idle: (1)~limited parallelism, (2)~high latency of two-qubit gates, and (3)~additional data movement due to SWAP operations. Quantum programs have limited operational parallelism as complex multi-qubit operations are translated into a highly serial sequence of two-qubit CNOT gates. Similarly, there exists significant non-uniformity in the latency of different operations on NISQ computers. For example, the latency of a CNOT gate on IBMQ hardware ($\approx$ 400 ns) is almost an order of magnitude higher than the latency of a single qubit gate. Also, CNOT gates on the same hardware incur different latencies. For example, while the latency of a CNOT gate is 440 ns on average, it can be as high as 860 ns on IBMQ-Toronto.  Therefore, even if a program can orchestrate parallel operations on different qubits, some with single-qubit operations and others with two-qubit operations, the qubits with single-qubit operations finish execution earlier than the qubits with two-qubit operations and remain idle. Similarly, parallel two-qubit gates with variable latencies finish execution at different times. Finally, architectural constraints can also cause idle times in programs as current machines do not have all-to-all connectivity.  When two unconnected qubits need to perform a two-qubit operation, the compiler inserts SWAP instructions (typically performed as a sequence of 3 CNOT instructions) to perform data movement, which causes serialization and large idle periods due to the long latency of these operations.

By keeping the idle qubits busy with a specific sequence of gates, their susceptibility to spurious noise can be reduced.
{\em Dynamical Decoupling (DD)}~\cite{oliver,DDLidarAPS,DDLidarPeriodic} is a well-known technique from the quantum devices community that uses this insight to mitigate idling errors. DD is implemented by the repeated execution of a sequence of single-qubit operations that return the qubit to its original state. Thus, DD operations do not change the overall state of the qubits as they collectively behave as an identity gate that suppresses noise.  DD is widely used in device-level characterization experiments to remove systematic noise~\cite{oliver}. More recently, DD has been adopted in quantum volume~\cite{IBMDD} and quantum error correction circuits~\cite{chen2021exponential}.

So far, the use of DD has mainly been limited to single qubit experiments or specific circuits~\cite{IBMDD,chen2021exponential}. Most importantly, at the application-level, prior studies have relied on enabling DD for all idle qubits. However, our experiments on IBMQ systems show that such indiscriminate application of DD often results in sub-optimal improvements in fidelity because while DD can reduce idling errors, it can still increase the effective error-rate of a qubit if the errors introduced due to the extra operations outweighs the benefits. 


Moreover, the effectiveness of DD depends on the structure of the program, qubit state, and device characteristics.  We explain this dependency with an example. Figure~\ref{fig:intro}(a) shows a quantum program with 3-qubits. Note the difference in instruction latencies (40ns for $H$ versus 400ns and 600ns for the $CNOT$s) and that qubits $\mathsf{q}[0]$ and $\mathsf{q}[2]$ remain idle for significant periods of time.  Figure~\ref{fig:intro}(b-d) shows the three options for applying DD -- to all idle qubits, only $\mathsf{q}[0]$, and only $\mathsf{q}[2]$, respectively. Figure~\ref{fig:intro}(e) compares the reliability of the baseline with the three options for DD.  While applying DD for all idle qubits provides some reliability benefits, the highest improvement comes when DD is applied to only a subset of the qubits (only $\mathsf{q}[2]$ in this example). Thus, to mitigate idling errors at the application-level, DD must be applied robustly and judiciously for the most optimal performance. To that end, this paper proposes {\em Adaptive Dynamical Decoupling (ADAPT)}, a software framework to reliably use DD for NISQ applications by identifying the subset of qubits that are most likely to benefit from DD. 

\begin{figure*}[h!]
\begin{centering}
    \includegraphics[width=\textwidth]{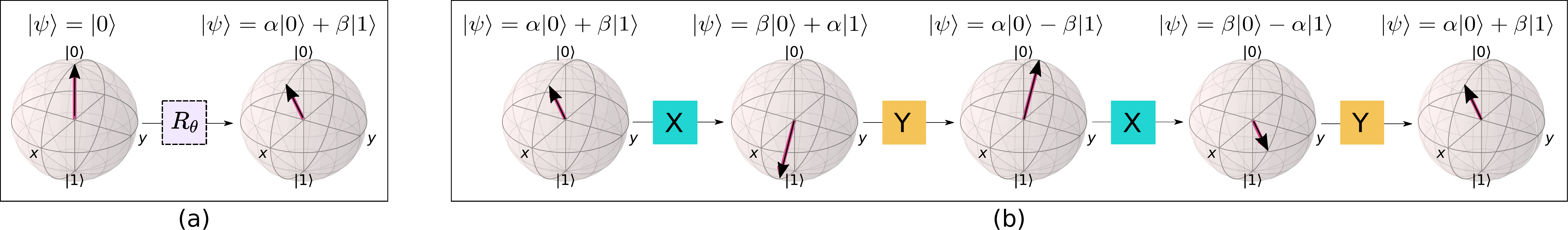}
    \caption{(a) Qubit-rotation operation on a single qubit  (b) Dynamical Decoupling (DD) using XYXY (XY4) sequence.}
        \label{fig:back}
\end{centering}

\end{figure*}

ADAPT employs a trial-and-error method to search for the subset of qubits that maximizes the application fidelity. However, the output of the program must be known a-priori for this search to be effective, which is not possible. We observe that the effectiveness of DD depends on the program structure and two programs with similar structures tend to have similar fidelity when executed on identical physical qubits. ADAPT leverages this insight and searches for the optimal DD sequence using a {\em Decoy Circuit} that is structurally similar to the given program but built using Clifford operations (and only a handful of non-Clifford instructions). Clifford gates (CNOT, H, X, Z, S) can be efficiently simulated on conventional computers~\cite{Knill} and therefore, the noise-free output of the decoy circuit can be estimated. The decoy circuit shows a similar trend in idling errors as the input circuit because it uses identical CNOT operations as the input circuit. ADAPT performs the trial-and-error search on the decoy circuit by inserting different DD sequences on the subset of the qubits and selecting the pattern that maximizes the likelihood of getting the correct answer on the decoy circuit. While the number of possible DD combinations increases exponentially with the number of qubits, ADAPT employs a divide-and-conquer approach where only 4 qubits are evaluated exhaustively at any time, keeping the search tractable. This results in a linear complexity in the number of qubits.

We evaluate the effectiveness of ADAPT using IBMQ systems, ranging from 16 to 27 qubits, and two different DD protocols. Our evaluations show that ADAPT is robust and improves fidelity for both types of DD sequences, making it generalizable to other systems and DD protocols. ADAPT improves the fidelity of key NISQ benchmarks on average by 1.86x and by up-to 5.73x compared to the baseline without DD, and by 1.2x compared to applying DD to all qubits. The software for ADAPT and datasets for the evaluations in this paper is available  at this \href{https://github.com/pdas36/ADAPT}{\underline{link}}.

\vspace{0.03in}

Overall, this paper makes the following contributions:
\begin{enumerate}[leftmargin=0cm,itemindent=.5cm,labelwidth=\itemindent,labelsep=0cm,align=left, itemsep=0.08cm, listparindent=0.3cm]

\item To the best of our knowledge, this is the first paper to evaluate DD at an application level.  We show that while DD is beneficial in general, applying DD indiscriminately to all the idle qubits does not provide the highest fidelity.

\item We propose ADAPT, a software framework that applies DD judiciously by estimating the subset of qubits that are likely to provide the highest reliability with DD.

\item We present a Clifford-based decoy circuit approach and an efficient search algorithm to practically implement ADAPT.

\end{enumerate}

\section{Background}
\subsection{Quantum Bits and Gates}
The state of a qubit is denoted as a superposition of its basis states $\ket{0}$ and $\ket{1}$, i.e., $\ket{\psi} = \alpha\ket{0} + \beta\ket{1}$. It can be represented as a point on the Bloch sphere, as shown in the Figure~\ref{fig:back}(a). When a qubit with state $\ket{\psi}$ is measured, it produces "0" with a probability of $\alpha^2$ and "1" with probability  $\beta^2$.  A single qubit gate rotates the qubit state from one point on the sphere to another, as shown in Figure~\ref{fig:back}(a).


\subsection{NISQ Hardware Errors}
\label{sec:nisqerrors}

For a qubit in superposition, even a tiny change in its energy can produce a valid but \textit{erroneous} state. The likelihood of undesirable changes in the state of a qubit is defined as the qubit error rate. Qubit errors can broadly be classified into:

\vspace{0.05 in}
\noindent{\textbf{Active Errors: }}These errors are caused when the qubit is performing gate or measurement operations.  On current generation of quantum computers from IBM, the error rate is approximately 0.1\% for single-qubit operations, 1\%-2\% for two-qubit operations, and 4\% for measurement operations.

\vspace{0.05 in}
\noindent{\textbf{Idling Errors: }}These errors occur when a qubit is idle and not performing any operations. For example, coherence errors can cause qubits to naturally decay to the lowest energy state ($\ket{0}$) within a short time ($10-100$ $\mu\textrm{seconds}$). Moreover, qubits can lose their phase information due to their interactions with the environment (dephasing).  Similarly, undesirable crosstalk can cause operations on other active qubits to affect the state of neighboring (spectator) idle qubits. Idling errors can also be caused by environmental noise. 

\ignore{
\begin{figure*}[t!]
\centering
    \includegraphics[width=2\columnwidth]{micro54-latex-template/Figures/Fig_Motivational_Exp1.pdf}
    \caption{(a) Circuit to evolve qubit q[0] freely (b) Circuit to evolve q[0] with DD (c) Fidelity of q[0] with free evolution and DD (d) Circuit to evolve q[0] in presence of cross-talk (e) Circuit to evolve q[0] with DD in presence of cross-talk (f) Fidelity of q[0] in presence of cross-talk, with and without DD}
    \vspace{-0.15in}
    \label{fig:motiv}
\end{figure*}

}

\subsection{NISQ Model for Quantum Computing}
Hardware errors can cause a quantum program to produce an incorrect answer.  Unfortunately, in the near term, quantum computers will not have enough resources to perform error correction (which can incur 20x-100x overhead). However, some quantum algorithms can tolerate limited hardware errors at the algorithmic level and can be used to solve practical problems using the {\em Noisy Intermediate-Scale Quantum (NISQ)} model, wherein the program is run thousands of times to identify the correct answer. NISQ systems can solve problems that are beyond the reach of existing computers~\cite{GoogleQ,preskillNISQ}. However, the ability to infer the correct answer on NISQ machines depend on the error-rates and program length.

Recent works have proposed software solutions to reduce the length of programs~\cite{li2018tackling,DACWille,wille2014optimal,CGO} and use error characteristics to improve program fidelity~\cite{tannu2019not,murali2019noise,murali2020software}. Although useful in mitigating measurement, gate errors, and CNOT-CNOT crosstalk errors~\cite{murali2020software}, these schemes do not focus on idling errors. In this paper, we focus on software policies to mitigate idling errors at  application-level. 

\subsection{Idling Errors in NISQ Applications}
\begin{figure}[ht]
\centering
    \includegraphics[width=\columnwidth]{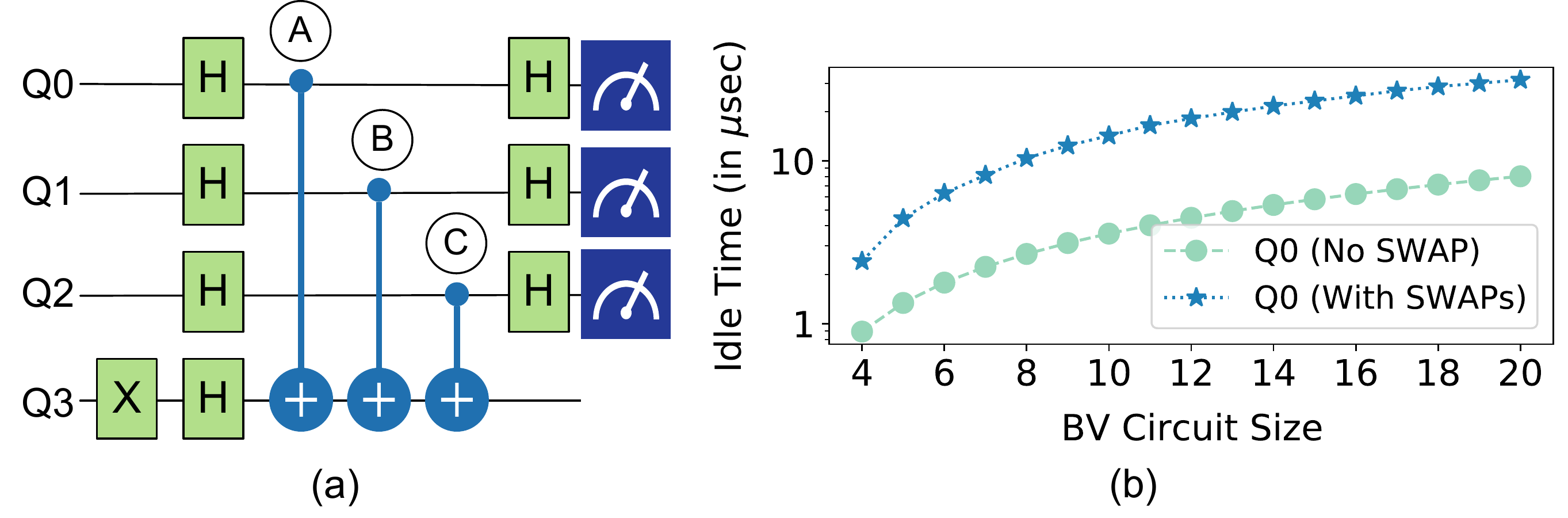}
    \caption{(a) 4-qubit Bernstein-Vazirani circuit (b) Impact of SWAPs on the idle time of Q0 when BV circuits with increasing sizes are executed on IBMQ-Toronto.} 
    \label{fig:bv_variable}
\end{figure}

The characteristics of quantum programs cause many of the qubits to remain idle for a significant period of time during program execution. There are three key reasons for qubits to remain idle: (1)~limited parallelism, as quantum programs get decomposed into sequence of instructions with data dependency (2)~high latency of two-qubit gates compared to single-qubit gates, and (3)~additional data movement or SWAP operations. For example, let us take a 4-qubit Bernstein-Vazirani (BV) circuit shown in Figure~\ref{fig:bv_variable}(a). Due to low parallelism, the CNOT gates of this circuit must be scheduled serially. Consequently, qubit Q0 remains idle when CNOTs \circled{B} and \circled{C} are executed. Although existing compilers minimize idle times by scheduling instructions \textit{as late as possible}~\cite{qiskit,murali2020software}, this optimization is not feasible for all qubits as the computation must make forward progress. For example, late initialization causes Q2 to not experience any idle time but cannot optimize Q1 from remaining idle during the execution of CNOT \circled{C}. The long latency of the CNOT gates exacerbates the idle times. Even if a program can orchestrate parallel operations on different qubits, qubits with single-qubit operations (10x faster) finish execution earlier than qubits with two-qubit gate operations and remain idle. Furthermore, CNOT gates on the same hardware incur different latencies. For example, the worst-case CNOT gate latency on IBMQ-Toronto is 1.95x the average latency. Consequently, parallel CNOT gates with variable latencies finish at different times. Finally, NISQ compilers insert SWAP instructions to overcome limited device connectivity causing serialization and long idle periods. For example, Figure~\ref{fig:bv_variable}(b) shows that the idle time of qubit Q0 for different BV circuits is about an order of magnitude higher on IBMQ-Toronto compared to a machine with similar error rates but all-to-all connectivity.

\begin{table}[htp]
\centering
\begin{small}
\caption{ Idling Times for Programs on IBMQ-Rome}
\setlength{\tabcolsep}{0.1cm} 
\renewcommand{\arraystretch}{1.2}
\begin{tabular}{ | c || c || c | c | c | c | c || c | c |}
\hline
   Workload & Program & \multicolumn{5}{c||}{Idle Fraction (\%)} & \multicolumn{2}{c|}{Fidelity} \\  
   \cline{3-9}
           Name & Latency & $Q_0$ & $Q_1$   & $Q_2$  &  $Q_3$  &  $Q_4$ & No DD  & DD on all\\ 
          
   \hline \hline 
   QFT-5 & $13.1\ \mu S$  & 92 &  38 & 17\ & 33\ & 63\ & 0.18 & 0.41 
  \\ \hline 
  
    QAOA-5  & $2.50\ \mu S$ & 82\ & 37\ & 35\ & 63\ & 79\ & 0.64 & 0.80
  \\ \hline 
  
 Adder & $9.90\ \mu S$ & 41 & 42\ & 9\ & 42\ & 75\ & 0.40 & 0.37
  \\ \hline 

\end{tabular}
\label{tab:idle}
\end{small}
\end{table}

Table~\ref{tab:idle} shows the program latency and the percentage of time for which each qubit in three different  five-qubit programs remains idle on IBMQ-Rome. Note that across these workloads, qubits remain idle on an average more than 50\% of the time, and as much as 92\%.

\begin{figure*}[t]
\centering
    \includegraphics[width=\textwidth]{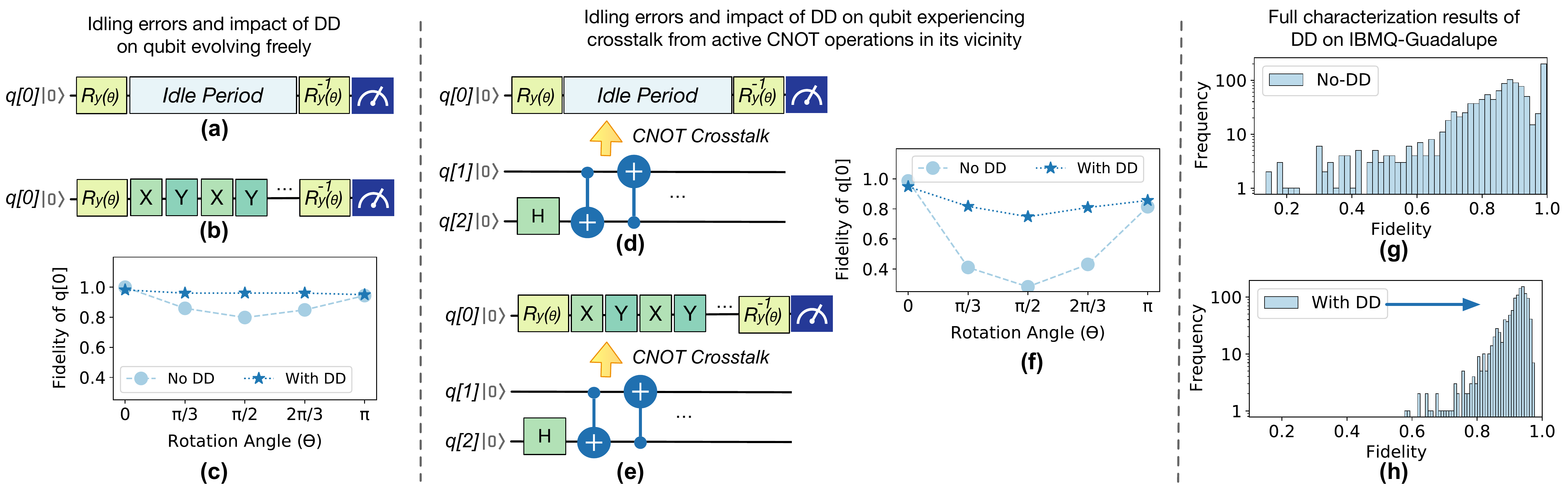}
    \caption{Circuit to evolve qubit q[0] (a) freely (b) with DD. (c) Fidelity of q[0] with free evolution and with DD. Circuit to evolve q[0] (d) freely (e) with DD in presence of crosstalk from on-going CNOT operations. (f) Fidelity of q[0] in the presence of crosstalk, with and without DD.  Distribution of fidelity of the idle qubit (g) without and (h) with DD when circuits (d-e) are executed on every qubit-link combination on IBMQ-Guadalupe.} 
    \label{fig:motiv}
\end{figure*}

\subsection{Dynamical Decoupling for Idling Errors}

To minimize the impact of idling errors, experimentalists have proposed {\em Dynamical Decoupling (DD)}~\cite{oliver,DDLidarAPS,DDLidarPeriodic}. As shown in Figure~\ref{fig:back}(b), DD keeps an idle qubit active by continuously rotating its state using single-qubit operations which suppresses the coupling between environmental noise and the qubit. DD is implemented by repeated execution of a sequence of single-qubit operations that returns the qubit to its original state (for example, a sequence of XYXY operations, as shown in Figure~\ref{fig:back}(b)). Thus, DD operations do not change the qubit's overall state as they collectively behave as an identity gate that suppresses the noise. DD is widely used in qubit characterization and has been shown to be effective on IBM~\cite{IBMDD}, Rigetti~\cite{pokharel2018demonstration}, and Google quantum hardware~\cite{chen2021exponential}.

\subsection{The Drawback of Dynamical Decoupling}
While DD can reduce idling errors, the extra operations introduced by it can cause gate errors. If the error-rate of these extra operations exceed the reduction in idling errors, then employing DD can reduce fidelity. Thus far, DD has been primarily limited to device-level studies for characterizing single qubits and its efficacy in reducing idling errors at the application-level is not yet fully understood.

Table~\ref{tab:idle} shows the application fidelity for the two cases (a)~when the program does not employ DD and (b)~when all qubits employ DD during the idle periods. We observe that applying DD on all the qubits can improve fidelity. However, our characterization studies show that even when DD improves application fidelity, we may obtain even higher fidelity by applying DD to only a select subset of qubits.  While DD has been an effective at the single qubit level, it is unclear how to apply robustly DD at the application level.
 
\subsection{Goal: Software for Robust Use of DD}
The goal of this paper is to develop a software framework that can enable robust use of DD at the application level. As idling errors are program specific, the subset of qubits that benefit from DD are unique to each program. Therefore, our framework must identify, for each program, the sequence of DD activations by taking the program structure into account. To this end, we propose {\em Adaptive Dynamical Decoupling (ADAPT)}, which tries to learn the subset of qubits that must use DD to obtain the highest fidelity for a given program. Before we describe our solution, we present some characterization data  and explain the challenges in applying DD at the application level to motivate the use of DD and our solution.

\section{Characterizing the Effectiveness of Dynamical Decoupling}
\label{sec:motivation}

\subsection{Mitigating Idling Errors using DD}
We quantify idling errors and the impact of DD in mitigating these errors using the characterization circuits shown in Figure~\ref{fig:motiv}(a) and (b). In the first circuit, shown in Figure~\ref{fig:motiv}(a), we initialize the qubit $\mathsf{q}[0]$ in an arbitrary state by rotating along the Y-axis, using an $R_y({\theta})$ gate, and allow it to evolve freely over the idle period. This is achieved on IBMQ systems by inserting Delay or Identity gates. In the end, we bring the qubit back to state $\ket{0}$ by performing an inverse rotation, $R_y^{-1}({\theta})$, and measure it. To understand if the errors are more than just natural decoherence, we perform a similar experiment. However, now we use XY pulses, typically used for DD, throughout the idle period, as shown in Figure~\ref{fig:motiv}(b). Consequently, qubit $\mathsf{q}[0]$ does not remain idle in this circuit. 
We prepare multiple initial states by choosing different values of $\theta$ and study the circuits for $1.2\ \mu s$ (typical latency of 3 CNOTs). Figure~\ref{fig:motiv}(c) shows the fidelity of this circuit for a qubit on IBMQ-London. We observe that the fidelity of qubit $\mathsf{q}[0]$ improves significantly when DD is applied.

\subsection{Effectiveness of DD under Crosstalk}
\label{sec:motivation2}
In quantum programs, a qubit typically remains idle when other qubits are actively performing gate operations. Prior studies show that on-going two-qubit CNOT operations generate crosstalk that can significantly lower the fidelity of concurrent CNOT operations~\cite{murali2020software,harper}. To study the impact of crosstalk from concurrent CNOT operations on idling errors, we characterize the fidelity of an idle qubit by executing the circuits shown in Figure~\ref{fig:motiv}(d-e). In the first circuit, qubit $\mathsf{q}[0]$ evolves freely in the presence of CNOT operations on neighboring qubits, whereas in the second circuit, the qubit $\mathsf{q}[0]$ evolves in the presence of the XY DD gate sequence. Running these circuits on IBMQ-London for five different quantum states of qubit $\mathsf{q}[0]$ and an idle time period of $2.4\ \mu s$, we observe that the fidelity of the idle qubit $\mathsf{q}[0]$ drops up-to 34\% in the presence of concurrent CNOT operations, as shown in Figure~\ref{fig:motiv}(f). Thus, idling errors get amplified significantly in the presence of crosstalk, making quantum programs extremely vulnerable to these errors. We also observe DD to be effective even in the presence of crosstalk from on-going operations, as the fidelity improves from 34\% to 75\%.

We also characterize idling errors on 16-qubit IBMQ-Guadalupe. We map the idle qubit $\mathsf{q}[0]$ to every physical qubit. Further, for each idle qubit, the active qubits -- $\mathsf{q}[1]$ and $\mathsf{q}[2]$, are mapped to any of the remaining fifteen qubits that are physically connected. On IBMQ-Guadalupe, there are 224 such possible combinations and for each qubit-link combination, we run the two circuits (without and with DD) for five different theta values (initialized using $\theta$ in $[0,\pi]$). We also increase the idle time to 8 $\mu s$ to understand idling errors and the effectiveness of DD in the context of large programs. Note that $8\ \mu s$ is reasonable, as even a small program with 6-8 qubits and 20 serial CNOTs can experience such large idle periods. Figure~\ref{fig:motiv}(g-h) shows the distribution of the fidelity of these 2240 ($= 224\times 5\times 2$) circuits, without and with DD respectively. We observe that the fidelity of the idle qubit drops to 84.5\% on an average and up-to 13.6\% in the worst-case, when it evolves freely. However, with DD, the average fidelity improves to 91.3\% and 57.7\% in the worst-case.

\subsection{Factors Influencing Idling Errors and DD}
\begin{figure}[htp]
\centering
    \includegraphics[width=1\columnwidth]{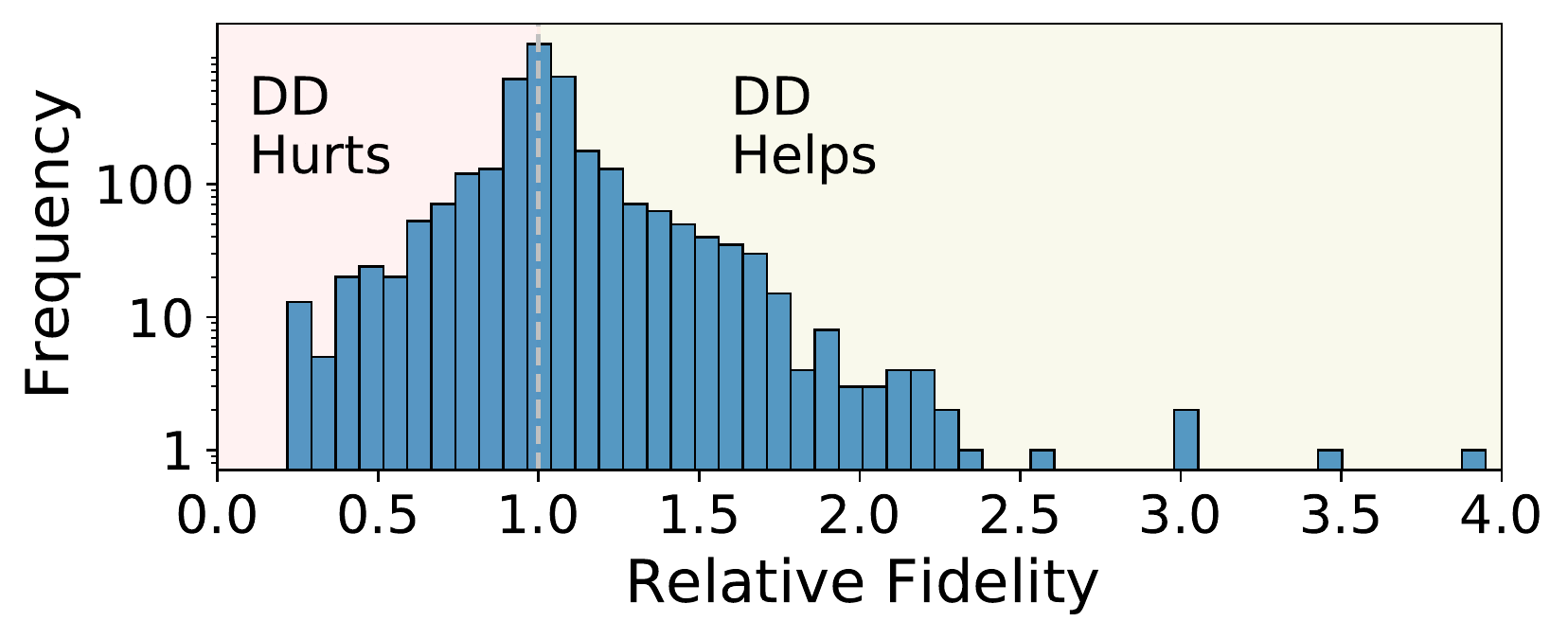}
    \caption{Distribution of Relative Fidelity of $\mathsf{q}[0]$ in presence of DD for 700 qubit-link combinations on IBMQ Toronto.}
    \label{fig:histogram}
\end{figure}

To understand factors influencing idling errors and the effectiveness of DD, we repeat the characterization experiments discussed in Section~\ref{sec:motivation2} on other IBMQ systems across multiple calibration cycles. For example, on 27-qubit IBMQ-Toronto, there are 700 qubit-link combinations and we characterize each of them using a total of 7000 circuits. We make the following key observations:

While DD generally improves the fidelity of the idle qubit $\mathsf{q}[0]$, there are many instances where DD worsens the fidelity. For example, Figure~\ref{fig:histogram} shows the histogram of the relative fidelity of qubit $\mathsf{q}[0]$ in the presence of DD. While DD improves the fidelity up-to 3.95x in the best case, it lowers the fidelity to 0.21x in the worst-case.

\begin{tcolorbox}
Applying DD pulses to every qubit during each idle time window can adversely impact the fidelity of a program.
\end{tcolorbox}

\begin{figure}[h]
\centering
    \includegraphics[width=0.85\columnwidth]{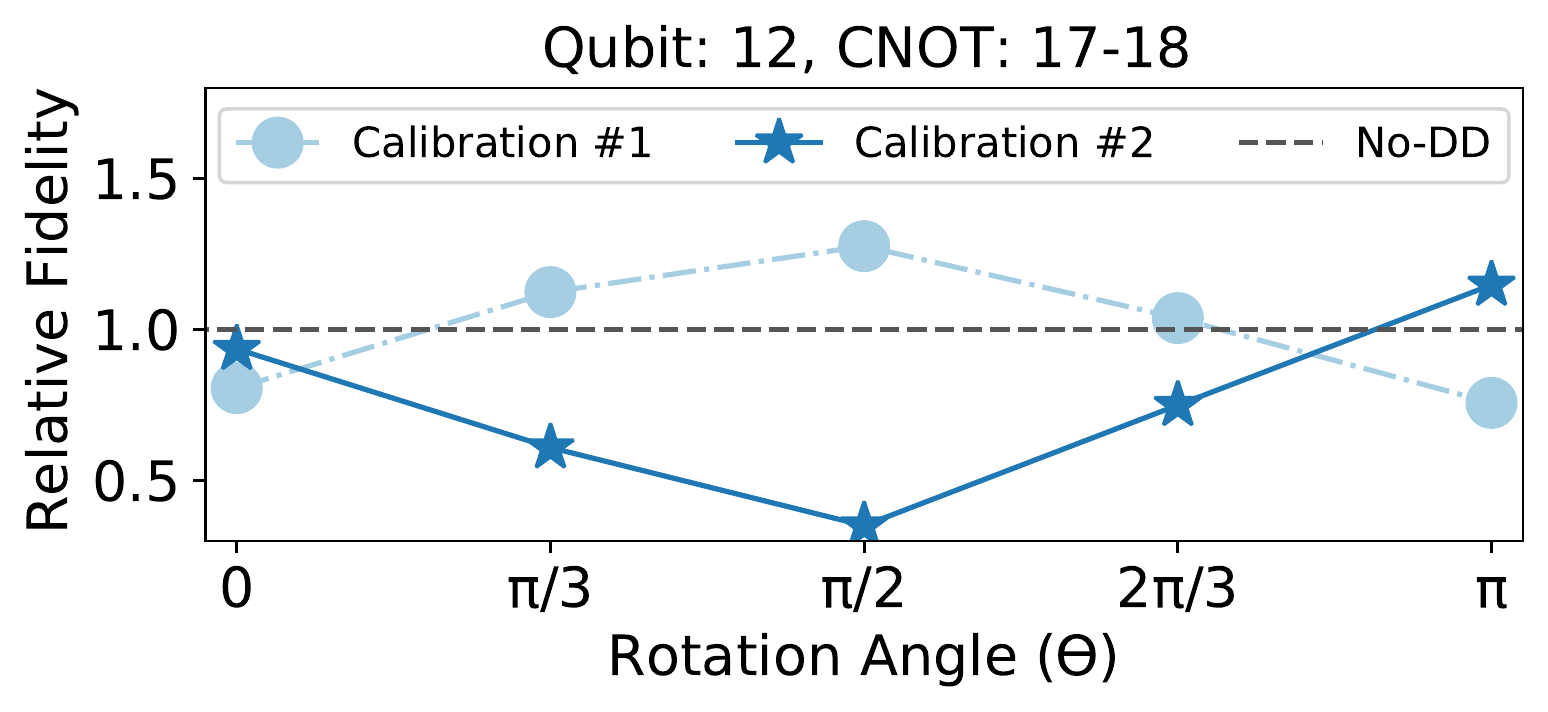}
    \caption{Relative Fidelity of Qubit-12 in the presence of  CNOTs on Link:17-18 for different calibration cycles.}
    \label{fig:downsideofdd}
\end{figure}

Idling errors depend on the state of qubits and concurrent CNOTs. Further, the effectiveness of DD change across calibration cycles. For example, Figure~\ref{fig:downsideofdd} shows the relative fidelity of Qubit-12 when CNOT operations are performed on the Link:17-18 for two different calibration cycles. In the first cycle, DD improves the fidelity up-to 1.27x, whereas DD degrades the fidelity up-to 0.35x in the second cycle. Our experiments also show that idling errors exist between qubit-link pairs that may not be present in the same on-chip neighborhood, making localized characterization approaches~\cite{murali2020software} inadequate. We make similar observations using other (a)~systems such as IBMQ-Paris and IBMQ-Casablanca and (b)~DD pulses~\cite{IBMDD}.

\begin{figure*}[t]
\centering
    \includegraphics[width=\textwidth]{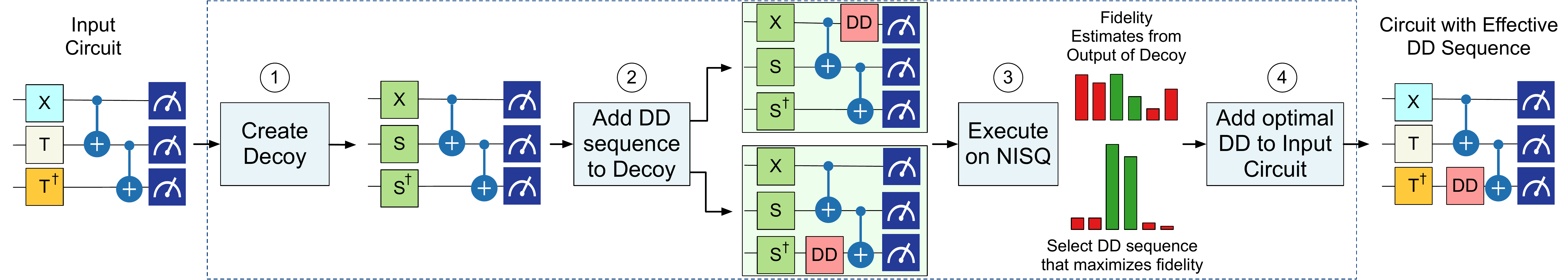}
    \caption{Overview of ADAPT with key building blocks} 
    \label{fig:overview}
\end{figure*}

\begin{tcolorbox}
It is impractical to estimate the effectiveness of DD for all possible quantum states with respect to each qubit-link combination in large systems every calibration cycle.
\end{tcolorbox}

The characterization complexity increases further when simultaneous CNOT operations are considered on multiple links. Our experiments show that although CNOT operations on multiple links can result in higher idling errors in general, such additive effect does not exist always. For some of the cases, one of the active links dictates the overall idling error. In rare situations, multiple CNOTs can reduce idling errors.

\vspace{0.06in}
To summarize, our evaluations using IBMQ systems show complex trends in idling errors and effectiveness of DD. We confirm that (1)~on-going CNOTs increase idling errors, (2)~the idling error-rate and effectiveness of DD depends on the CNOT patterns and combination of the idle and active qubits, and (3)~the idle duration.

\subsection{Impact of DD on Application Fidelity}
The most straightforward method to enable dynamical decoupling at the application level is to insert DD pulses wherever feasible. To implement this design, we can identify all program regions where each qubit is idle and insert DD sequences. However, naively inserting DD sequences for all qubits may not always be beneficial, as we have already observed from the characterization experiments. To demonstrate this effect at the application-level, we execute two 6-qubit benchmarks, {\em Quantum Fourier Transform (QFT)} and {\em Bernstein Vazirani (BV)}, on 27-qubit IBMQ-Toronto, with DD applied on all 64 ($2^6$) possible qubit combinations.  

\begin{figure}[htp]
    \centering 
    \includegraphics[width=\columnwidth]{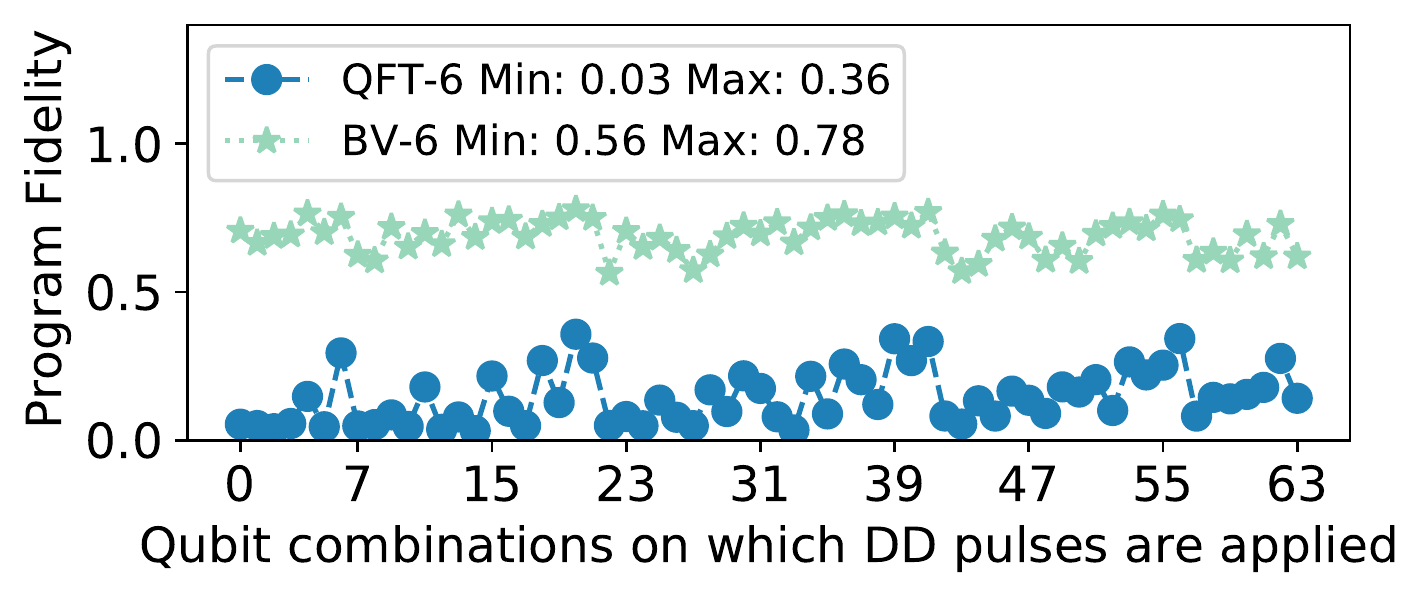}
    \caption{Fidelity of QFT and BV benchmarks with all possible DD sequences on IBMQ-Toronto. The sequence 0 \texttt{(000000)} denotes DD on none of the qubits and sequence 63 \texttt{(111111)} denotes DD on all qubits. \ignore{Note that the DD sequence with highest fidelity differs from the two extremes.}} 
    \label{fig:qft_adder_all_dd}
\end{figure}

Figure~\ref{fig:qft_adder_all_dd} shows the fidelity (likelihood of getting correct answer) for all 64 DD combinations, with 0 (000000) being the baseline when no DD is applied, and 63 (111111) being the case when DD is applied on all six qubits. We observe that both benchmarks show significant variation in fidelity for different DD sequences. For \textit{QFT}, enabling DD for all qubits increases the fidelity by 2.6x, however, the fidelity can be improved up-to 6.6x by choosing sequence "010100". For \textit{BV}, applying DD on all qubits degrades fidelity to 0.88x, and we may deem DD to be counter-effective for this benchmark. However, using the DD sequence "010100" can improve the fidelity by 1.1x compared to no-DD and up-to 1.26x compared to DD-for-all. Note that the best DD sequence depends on the workload characteristics and the physical qubits used to run the program.  

\begin{tcolorbox}
Applying DD on all qubits can reduce program fidelity. To effectively mitigate idling errors, DD should be applied judiciously only to the qubits that benefit from DD in a quantum circuit.
\end{tcolorbox}
\newpage
\section{Adaptive Dynamical Decoupling }

To enable the robust use of DD at the application level, we propose {\em Adaptive Dynamical Decoupling (ADAPT)}. ADAPT identifies the combination of DD sequences that suppress idling errors and maximizes the application fidelity. ADAPT is implemented as a compiler pass that can be easily integrated with existing and future quantum compiler tool flows. In this section, we provide an overview of ADAPT, discuss the design issues in estimating the optimal DD sequence, propose scalable search algorithms for the same.

 \subsection{Overview of ADAPT}
ADAPT identifies all idle qubit slots in a quantum circuit and applies DD gate sequences during these idle periods. However, the most optimal subset of qubits on which DD must be applied is neither known a-priori to program execution nor practically feasible to obtain through extensive device characterization. To overcome this challenge, ADAPT relies on a \textit{Decoy Circuit} which is structurally similar to the input program, but with a known solution. Furthermore, to limit the complexity of the search for the optimal DD sequence, ADAPT employs a localized algorithm. Figure~\ref{fig:overview} shows an overview of ADAPT which accepts a quantum circuit as the input and outputs the circuit with the most optimal DD sequence. We discuss the specific design details of ADAPT next.

\subsection{Clifford Decoy Circuits (CDC)}
If the outcome of a quantum circuit is known, we can apply DD on different subsets of qubits and assess their effectiveness in improving fidelity.  Unfortunately, practical applications are hard to simulate using conventional computers, and their correct outcome is unknown. To overcome this challenge in estimating the optimal DD sequence, we leverage the following insights. 

\vspace{0.05 in}
\noindent \textit{Insight \#1: Not all quantum circuits are hard to simulate and circuits comprising of Clifford gates only can be simulated efficiently on conventional computers~\cite{Knill,Aaronson_2004}.}

\vspace{0.05 in}
\noindent \textit{Insight \#2: Our characterization experiments show that crosstalk from CNOT operations is a dominant source of idling errors. Thus, two circuits with similar CNOT structures encounter similar idling errors.}

\vspace{0.05 in}
ADAPT uses these two insights \noindent \circled{1}~to generate an efficient {\em Clifford Decoy Circuit (CDC)} that preserves the structure of the input program. \noindent \circled{2}~Next, ADAPT applies different DD combinations to this decoy circuit and \noindent \circled{3}~selects the DD sequence that maximizes the fidelity of the decoy circuit. \noindent \circled{4}~Finally, ADAPT applies this optimal DD sequence to the input circuit and executes it.

\subsubsection{Design of Clifford Decoy Circuits (CDC)} \hfill \\
\label{sec:cliffordreplacement}

\noindent ADAPT relies on Clifford Decoy Circuits generated using gates from the Clifford group -- {\em CNOT, X, Y, Z, H, S}. To create the CDC of a circuit, ADAPT replaces the non-Clifford gates of the circuit using the closest Clifford gates. 
To measure the closeness of a non-Clifford gate in the program with a Clifford gate, we use an operator norm- a distance measure used in the literature for approximating one unitary with another.    
\begin{align}
\|U-V\|_{\infty}:=\max _{|\psi\rangle \neq 0} \frac{\|(U-V)|\psi\rangle \|_{2}}{\||\psi\rangle \|_{2}}
\end{align}
For example,  by using operator norm, the U1 gate is either replaced by Z or S gates, whereas U2 and U3 gates are replaced by the closest Clifford gates depending on the Euler angles associated with the gates. As CNOT is a Clifford gate, the structure and usage of these gates are identical between the CDC and the input program and therefore, the CDC encounters similar crosstalk from CNOT operations. This also ensures that the qubits in the CDC experience similar idle times as the original circuit.

\color{black}
\subsubsection{Effectiveness of Clifford Decoy Circuits} 
\begin{figure}[htb]
    \centering 
    \includegraphics[width=1\columnwidth]{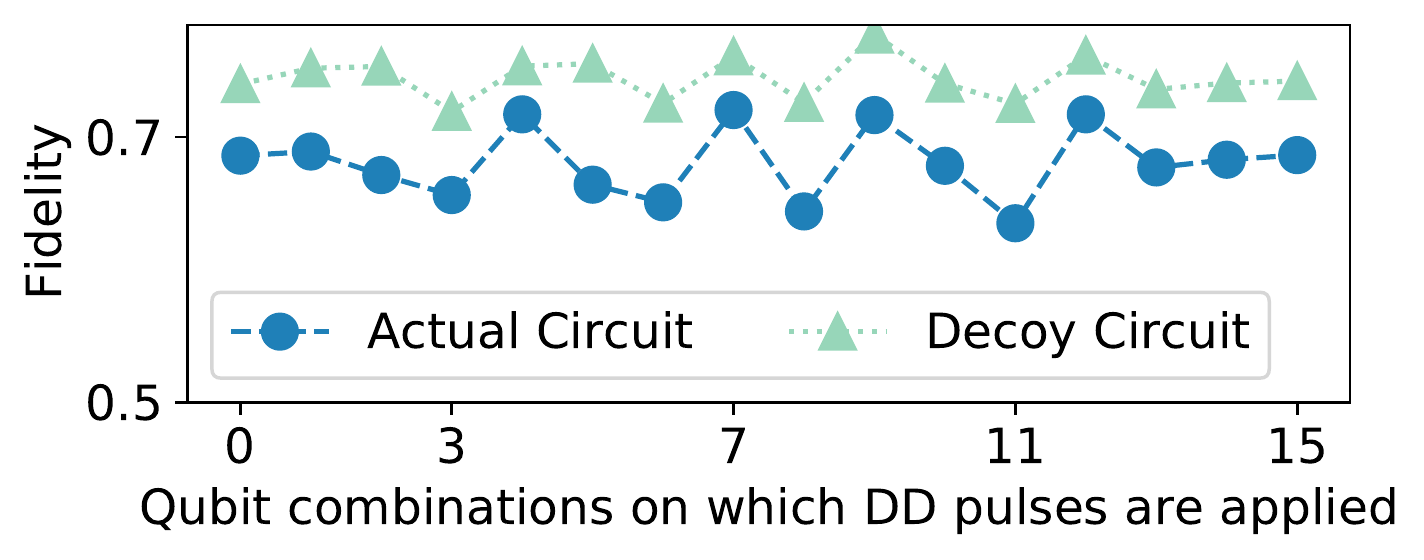}
    \caption{Correlation between the Fidelity of a 4-qubit Adder circuit on IBMQ-Guadalupe and corresponding Clifford decoy circuit. We observe strong correlation.} 
    \label{fig:adder_decoy_correlation}
\end{figure}

\hfill \\
\noindent To test the effectiveness of decoy circuits, we compare the fidelity of a 4-qubit quantum ADDER benchmark and its corresponding CDC for all possible DD sequence combinations. For example, this five qubit program has 16 ($2^4$) possible DD sequence combinations where the combination ``0 (0000)" indicates that DD is not applied on any of the qubits,  whereas the combination ``15 (1111)" indicates that DD is applied to all of the qubits.  We also compute the Spearman's Correlation Coefficient to quantify the agreement between the input program and the CDC. Figure~\ref{fig:adder_decoy_correlation} shows the trend in program fidelity for the actual circuit and the CDC and we observe that the program fidelity is strongly correlated to the fidelity of the CDC (Spearman's Correlation Coefficient = 0.78)

\subsubsection{Overcoming the Limitations of  CDCs: Seeded Decoy Circuits} \hfill \\
ADAPT generates sub-optimal DD sequences when there is a mismatch between the fidelity trend of the input program and its CDC. Our experiments show that a CDC with high variance in the output distribution can be insensitive to changes in idling errors and relying on them may result in sub-optimal DD sequences. For example, if a CDC produces a uniform distribution, executing this CDC with different DD sequences does not significantly change the output distribution in the presence of idling errors.

\begin{figure}[htb]
\centering
    \includegraphics[width=0.99\columnwidth]{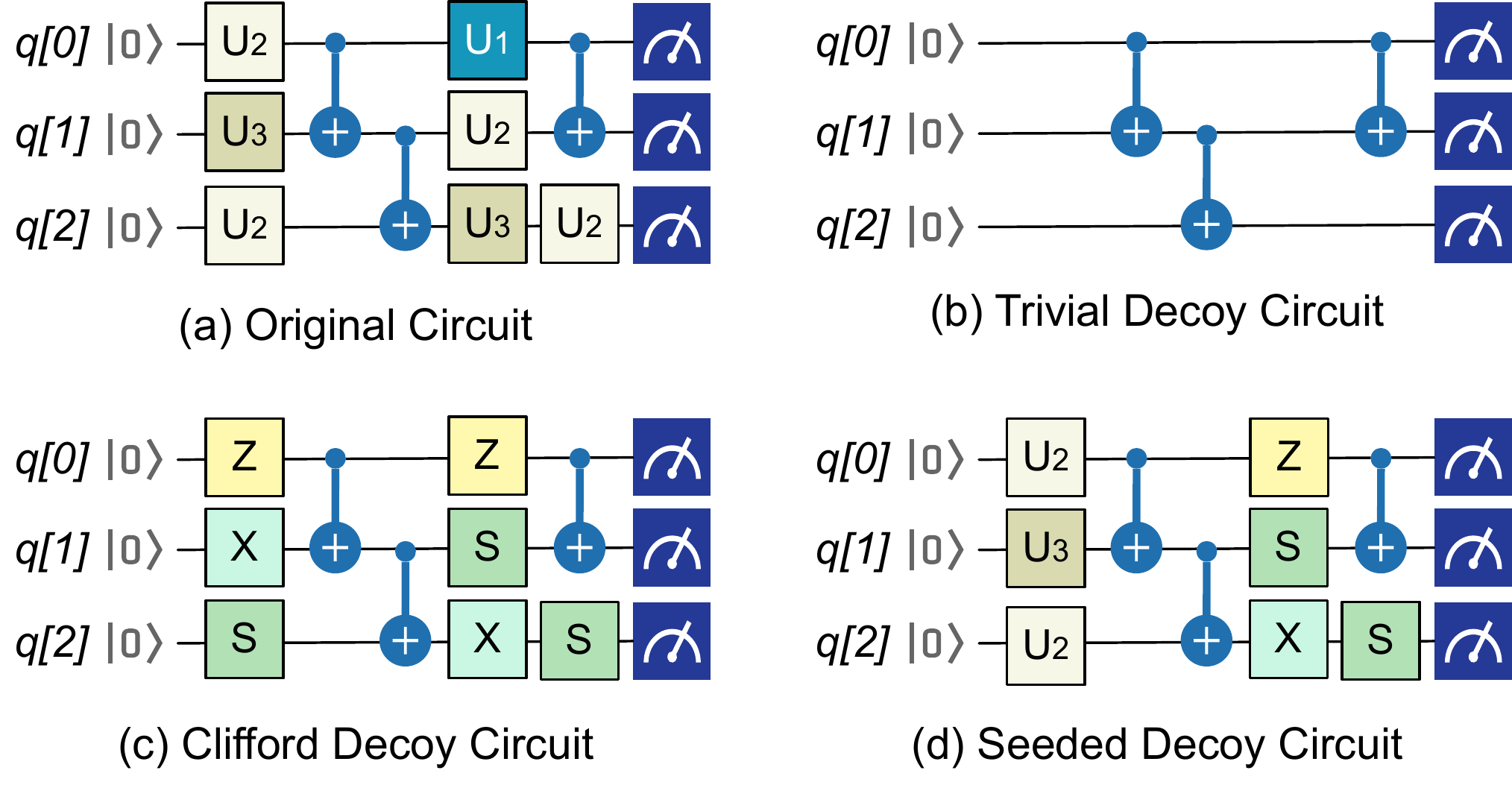}
    \caption{(a) Example quantum circuit. Decoy circuit construction using (b) Clifford gates (c) only CNOT (d) Mostly Clifford and few non-Clifford gates.} 
    \label{fig:sdc}
\end{figure}

To tackle this problem, we propose the use of {\em Seeded Clifford Decoy Circuits (SDC)} that generate output distributions with low entropy, thus making them sensitive to idling errors. While simply removing all the single qubit gates from an input program and preserving the CNOT structure only, as shown in Figure~\ref{fig:sdc}, can generate a decoy circuit, it does not truly mirror the fidelity trends of the input circuit because it does not capture the phase errors. To ensure that the output entropy is reduced while still being representative of the input circuit, SDCs use a very limited number of non-Clifford gates. SDCs apply an initial layer of non-Clifford gates on few qubits and replace the remaining non-Clifford gates with Clifford gates. For example, unlike the CDC shown in Figure~\ref{fig:sdc}(c) that uses all Clifford gates, the SDC shown in Figure~\ref{fig:sdc}(d) uses non-Clifford gates in the first layer of the circuit and Clifford gates in the later circuit layers. 

Our experiments show that SDCs can produce a rich state evolution while generating low entropy outputs. Although, the simulation cost of SDC is slightly higher than CDC, it is still not as expensive as running a full quantum simulation, which requires exponential resources. Moreover, SDCs produce low entropy outputs and thus, further optimizations to reduce the sampling cost can be deployed as well. For the purpose of simulations, we use Qiskit Extended Stabilizer Simulator (based on ~\cite{bravyi2019simulation}). Table~\ref{tab:SDC} shows the effectiveness of SDCs in improving the correlation between the decoy and original circuits and the time required to simulate the SDCs for 64,000 shots. To test the scalability, we simulate a 100-qubit QAOA SDC, which requires 330 seconds for 100,000 shots. Note that CDCs or SDCs require simulation only once because applying different DD sequences does not alter the output on a simulator. 

\begin{table}[htb]
\centering
\begin{small}
\caption{Correlation between Decoy and Input Circuits (higher is better)}
\setlength{\tabcolsep}{0.05cm} 
\renewcommand{\arraystretch}{1.2}
\begin{tabular}{ | c | c | c | c | c |  }
\hline
  Benchmark & Adder  & QFT-6  & QAOA-8 & QAOA-10         \\  \hline 
    Platform & IBMQ-Rome  & IBMQ-Paris & IBMQ-Paris & IBMQ-Paris         \\  \hline \hline 

  \textbf{CDC-Correlation} & 0.76  & 0.53 & 0.22 & -0.012         \\  \hline     
  \textbf{SDC-Correlation} & 0.81  & 0.68 & 0.74 & 0.62         \\  \hline \hline
  \textbf{SDC-SimTime} & 1.2 Sec  & 5.4 Sec & 12.8 Sec & 22.2 Sec         \\  \hline

\end{tabular}
\label{tab:SDC}
\end{small}

\end{table}

\begin{figure*}[htb]
\centering
    \includegraphics[width=0.7\textwidth]{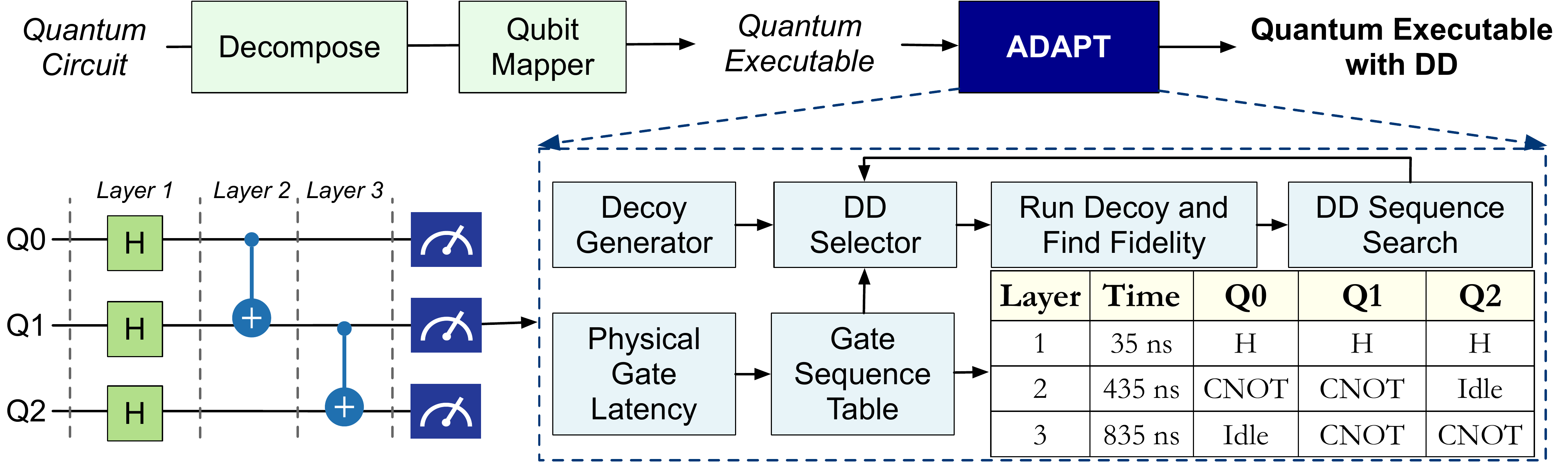}
    \caption{Overall Workflow of ADAPT} 
    \label{fig:adapt_schematic}
\end{figure*}

\newpage
\subsection{Managing Search Complexity}

The state-space for all possible DD sequences scales exponentially with the problem size. For a program with N qubits, there are $2^{N}$ possible combinations of qubits on which DD pulses can be applied when the qubits are idle. The combination $``000..0_{N}"$  represents DD applied on none of the qubits, whereas the combination $``111..1_{N}"$ represents DD applied on all the qubits, whenever the qubit is idle.  Unfortunately, we cannot search this design space exhaustively for large programs even with CDCs.  Instead, we perform localized search, whereby we first try to find the best subset of qubits in a neighborhood of 4-qubits (searching for all 16 combinations) before moving to the next neighborhood of 4-qubits. Therefore, for a circuit with N qubits, we use at most $4 \times N$ decoy circuits to estimate the best DD sequence for the input circuit. Thus, the search complexity increases linearly with the number of qubits. To accommodate the limitations of decoy circuits and obtain a more optimal DD sequence, we take a conservative estimate from the top two predicted sequences from ADAPT. For example, if the two best predictions are "1001" and "1011", the chosen sequence is "1011".

\subsection{Design Implementation}
Figure~\ref{fig:adapt_schematic} shows the workflow of ADAPT and how it fits into the existing NISQ execution model.
\subsubsection{Integration in the Compiler Tool-Flow} \hfill \\
ADAPT is applied at the end of existing compiler passes, after a program has been compiled into the machine-specific instructions of a given hardware. Typically, quantum compilers decompose a program into two-qubit CNOT and single qubit gates in the first pass and then map the program qubits on the physical devices. During both these compiler passes, redundant gates are eliminated. Moreover, the mapping pass schedules the gates to maximize the operational concurrency and minimize the number of SWAPs~\cite{li2018tackling}. Additionally, the mapping pass also accounts for the error-rates of the qubits and gates during qubit mapping and instruction scheduling~\cite{murali2019noise,tannu2019not}. ADAPT is orthogonal to these existing optimizations, and it is applied after all the compiler passes. Therefore, it can be seamlessly integrated with any existing NISQ compiler.

\subsubsection{Finding Idle Qubits} \hfill \\
To find idle qubits in a program, we translate the executable obtained from the compiler into an intermediate representation termed as the {\em Gate Sequence Table (GST)}, as shown in Figure~\ref{fig:adapt_schematic}. The GST slices the compiled circuit into layers and captures the data dependencies between the qubits in time. Note that the typical circuit representation used by the decomposition and mapping passes do not capture the idle cycles as gate latencies are not embedded in circuit representations. Whereas, GST uses timestamps generated by using physical gate latencies available from the machine calibration data to indicate start and end times of each gate. By querying the GST, we can identify the exact idle period for any qubit in a quantum program and insert the DD gate sequences. 

\subsubsection{Pulses for Implementing DD} 
\begin{figure}[htb]
\centering
    \includegraphics[width=1\columnwidth]{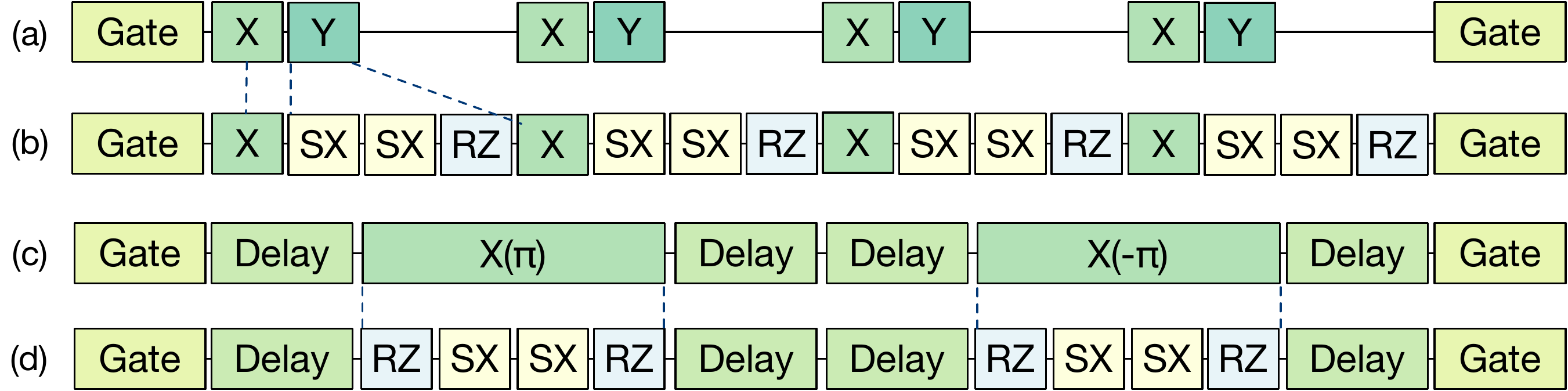}
    \caption{(a) XY-4 DD sequence, (b) decomposition of XY-4 sequence on IBMQ systems, (c) IBMQ-DD using X($\pi$)-- X($-\pi$) gates, and (d) decomposition of IBMQ-DD sequence inserted in the idle window between two gates.} 
    \label{fig:dd_sequences}
\end{figure}

\hfill \\
In theory, any sequence of instructions that are effectively identity gates can be used to implement DD. For example, we could simply use XX and YY pulses as the pairs result in identity operations. In this paper, we study two different DD protocols: the XY-4 sequence and the IBMQ-DD sequence as both have been shown to be effective for superconducting qubit devices in general~\cite{pokharel2018demonstration,chen2021exponential}, and particularly on IBMQ systems~\cite{pokharel2018demonstration,IBMDD}. The XY-4 sequence, shown in Figure~\ref{fig:dd_sequences}(a), is repeated execution of  "X-Y-X-Y"  gates, which takes about $210\ ns$ on most IBMQ machines as per its most optimal decomposition shown in Figure~\ref{fig:dd_sequences}(b). Note that since ADAPT adds DD sequences to the already compiled executable, it is necessary to add DD gates in the machine compliant instruction format. Each "X" and "SX" gate takes about $35\ ns$ and the "RZ" gate is performed in software~\cite{mckay2017efficient}. For this protocol, ADAPT inserts the DD gates for any idle windows with duration larger than $210\ ns$. Also, ADAPT continuously inserts XY-4 DD sequences for larger idle windows. For the IBMQ-DD sequence, ADAPT uses an approach similar to prior work~\cite{IBMDD} and inserts the decomposition of "X($\pi$)" and "X($-\pi$)" gates evenly during the idle period, as shown in Figure~\ref{fig:dd_sequences}. We also use a 10 nanosecond free evaluation buffer after each X and Y gate, consistent with the prior study on IBM systems~\cite{pokharel2018demonstration}.

\section{Experimental Methodology}
\label{sec:evaluation} 

In this section, we discuss the evaluation infrastructure used to the estimate the effectiveness of ADAPT. 
\subsection{Compiler}
We use IBM's Qiskit tool-chain to compile the benchmarks~\cite{qiskit}. We use the Qiskit transpiler with "{\em noise adaptive}" mapping~\cite{murali2019noise,tannu2018case}, "{\em sabre routing}" policy~\cite{li2018tackling}, and optimization level 3 flags. We ensure identical mapping and sequence of CNOT gate operations across all the policies evaluated for each benchmark. While we use this compiler tool-chain, ADAPT is independent of the compiler being used and therefore, any other compiler may be used as well. As ADAPT integrates with existing compilers as a post-compile step, we add the most optimal decomposition of the DD gates to obtain the final compiled quantum object. For the implementation of the DD pulses, we use both XY-4 and IBMQ-DD (XX) sequences because they have been previously studied for IBMQ systems and summarize the broad category of DD protocols available. Although beyond the scope of this paper, ADAPT can be implemented using other DD sequences and specialized gate pulses as well.

\subsection{Quantum Hardware Platforms}

For all our evaluations, we use three different quantum hardware from IBM. The details of each hardware and their average error trends are listed in Table~\ref{tab:ibmqplatforms}. Also, we perform all our evaluations for each benchmark and a quantum hardware within the same calibration cycle to establish a fair comparison. 

\begin{table}[htp]
\centering
\begin{small}
\caption{Error Characteristics of IBMQ Hardware}
\setlength{\tabcolsep}{0.15cm} 
\renewcommand{\arraystretch}{1.3}
\begin{tabular}{ | c || c || c | c || c | c | }
\hline
    \multirow{2}{*}{Machine Name} & Num. of & \multicolumn{2}{c|}{Error Rate (in \%)} & T1 & T2\\
    \cline{3-4} 
     & Qubits & CNOT & Measurement & ($\mu s$) & ($\mu s$) \\
    \hline \hline
    \texttt{IBMQ-Guadalupe} & {16} & 1.27 & 1.86 & 71.7 & 85.5 \\ \hline
    \texttt{IBMQ-Paris} & {27} & 1.28 & 2.47 & 80.8 & 83.4 \\ \hline
    \texttt{IBMQ-Toronto} & {27} & 1.52 & 4.42 & 105 & 114 \\ \hline
\end{tabular}
\label{tab:ibmqplatforms}
\end{small}
\end{table}

\ignore{
We also describe the latencies of the CNOT operations on each hardware in Table~\ref{tab:latency} as it dictates overall idle times experienced by a program in general. 
\begin{table}[htp]
\centering
\begin{small}
\vspace{-0.15 in}
\caption{Operational Latency on IBMQ Hardware}
\setlength{\tabcolsep}{0.15cm} 
\renewcommand{\arraystretch}{1.3}
\begin{tabular}{ | c || c || c | c | c|}
\hline
    \multirow{2}{*}{Machine Name} & Num. of &  \multicolumn{3}{c|}{CNOT Latency (in ns)}\\
    \cline{3-5} 
     & Physical Links & Min & Mean & Max \\
    \hline \hline
    \texttt{IBMQ-Guadalupe} & {16}& 263 & 394 & 583 \\ \hline
    \texttt{IBMQ-Paris} & {28} & 263 & 431 & 761 \\ \hline
    \texttt{IBMQ-Toronto} & {28} & 242 & 441 & 860 \\ \hline
\end{tabular}
\label{tab:latency}
\end{small}
\end{table}

}

\subsection{Benchmarks}
We use benchmarks of different sizes and structures to test the effectiveness of ADAPT. Table~\ref{tab:benchmarks} summarizes the benchmarks used in this paper. The size and type of benchmarks are derived from prior works on software mitigation of hardware errors~\cite{li2018tackling,tannu2019not,murali2019noise,micro1,nishio,li2020qasmbench}. Additionally, we run some of these benchmarks with different initial conditions. For example, QFT-7A and QFT-7B have identical structures, but compute the Fourier transform of two different quantum states. This tests the effectiveness of decoy circuits for the evolution of different quantum states.

\begin{table}[htp]
\centering
\begin{small}
\vspace{0.1 in}
\caption{ Quantum Benchmark Characteristics}
\vspace{0.1 in}
\setlength{\tabcolsep}{0.1cm} 
\renewcommand{\arraystretch}{1.3}
\begin{tabular}{ | c || c || c | c | c | c | }
\hline
    Benchmark & Benchmark & Num   & Total & Circ & Avg. Idle\\
    Description & Name & Qubits & Gates & Depth & Time ($\mu s$) \\
    \hline \hline
    \multirow{2}{*}{Bernstein Vazirani} & BV-7 & 7 & 95 & 30 & 8.8 \\
    \cline{2-6}
     & BV-8 & 8 & 132 & 55 & 15.2 \\
    \hline
    \hline 
    \multirow{4}{*}{Fourier Transform} & QFT-6A & 6 & 102 & 47 & 14.6 \\
    \cline{2-6}
     & QFT-6B & 6 & 228 & 148 & 27.9 \\
     \cline{2-6}
     & QFT-7A & 7 & 360& 202&50.0 \\
     \cline{2-6}
     & QFT-7B & 7 & 354 & 207 & 49.7 \\
    \hline
    \hline
    \multirow{4}{*}{Approx. Optimization} 
     & QAOA-8A & 8 & 32 & 19 & 4.6 \\
    \cline{2-6}
     & QAOA-8B & 8 & 60 & 30 & 8.1 \\
     \cline{2-6}
     & QAOA-10A & 10 & 77 & 37 & 13.5 \\
     \cline{2-6}
     & QAOA-10B & 10 & 180 & 50 & 7.7 \\
    \hline
    \hline
    Phase Estimation & QPEA-5 & 5 & 175 & 94 & 11.9 \\
    \hline
    
\end{tabular}
\label{tab:benchmarks}
\end{small}
\end{table}

\subsection{ Reliability Metrics}
To quantify the program reliability, we compare the output obtained on a real machine with respect to an idealized (error-free) machine (results obtained on an error-free simulator).  We quantify the program reliability by computing program Fidelity based on the distance between probability distributions. We use {\em Total Variation Distance (TVD)}~\cite{tvd:wiki}  to evaluate the fidelity by measuring the distance between ideal output probability (P) distribution and the real experiment's output (Q). A Fidelity of 1 means identical distributions, whereas 0 means completely different distributions. Therefore, a higher Fidelity is desirable.
\begin{center}
\vspace{-0.1in}
\begin{align}
{TVD\,(P, Q) = \frac{1}{2} \, \sum \|P_i-Q_i\| }
\end{align}
\end{center}

\begin{center}
\vspace{-0.1in}
\begin{align}
{Fidelity \, = \, 1-TVD \,(P, Q)}
\end{align}
\end{center}

Prior works have used similar metrics such as Success Probability~\cite{tannu2018case,murali2019noise,nishio}. We use distance-based metric as the output of the quantum program can be a probability distribution with multiple correct answers. Also, TVD closely matches prior metrics.

\subsection{Number of Trials}
We perform experiments with up-to 32,000 shots, depending on the program size, to obtain the output probability distributions. The largest benchmark in our study uses ten qubits (QAOA-10), so it can produce a maximum of 1024 ($2^{10}$) unique solutions and therefore, the number of samples used in our study are sufficiently large for the workloads.


\begin{figure*}[htp]
\centering
    \includegraphics[width=\linewidth]{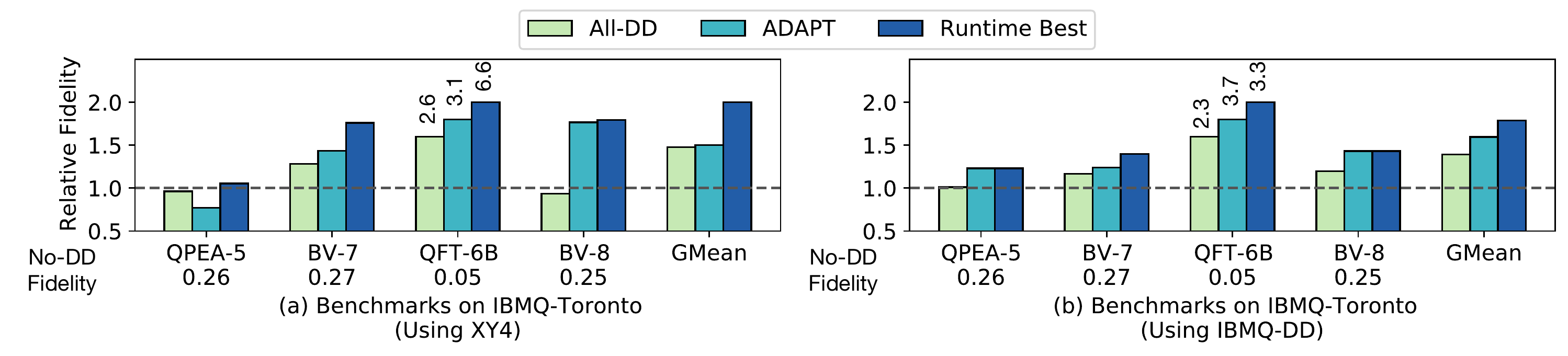}
    \caption{Relative Fidelity of different policies on 27-qubit IBMQ-Toronto using (a) XY4 and (b) IBMQ-DD sequences.}
    \label{fig:toronto_result}
\end{figure*}

\subsection{Competing Policies}

 For our evaluations, we use four competing policies which are described next:
 \vspace{0.05in}

\begin{enumerate}[leftmargin=0cm,itemindent=.5cm,labelwidth=\itemindent,labelsep=0cm,align=left, itemsep=0.08cm, listparindent=0.3cm]
    \item \textbf{No DD (Baseline)}: DD is not applied to any idle qubit.
    \item \textbf{All-DD}: DD is applied on all the program qubits during any time period when they are idle.
    \item \textbf{ADAPT}: The optimal DD sequence is obtained from a structured search using decoy circuits and applied.
    \item \textbf{Runtime Best}: Evaluates the program with all possible DD sequences ($2^N$ for an $N$-qubit program) and the sequence with highest fidelity at runtime is selected. 
\end{enumerate}

\section{Results}
In this section, we provide the evaluation results for ADAPT across three different quantum computers: 27-qubit IBMQ-Paris, 27-qubit IBMQ-Toronto, and 16-qubit IBMQ-Guadalupe. 

\subsection{Results for IBMQ-Paris}
\begin{figure}[htb]
\centering
    \includegraphics[width=0.95\columnwidth]{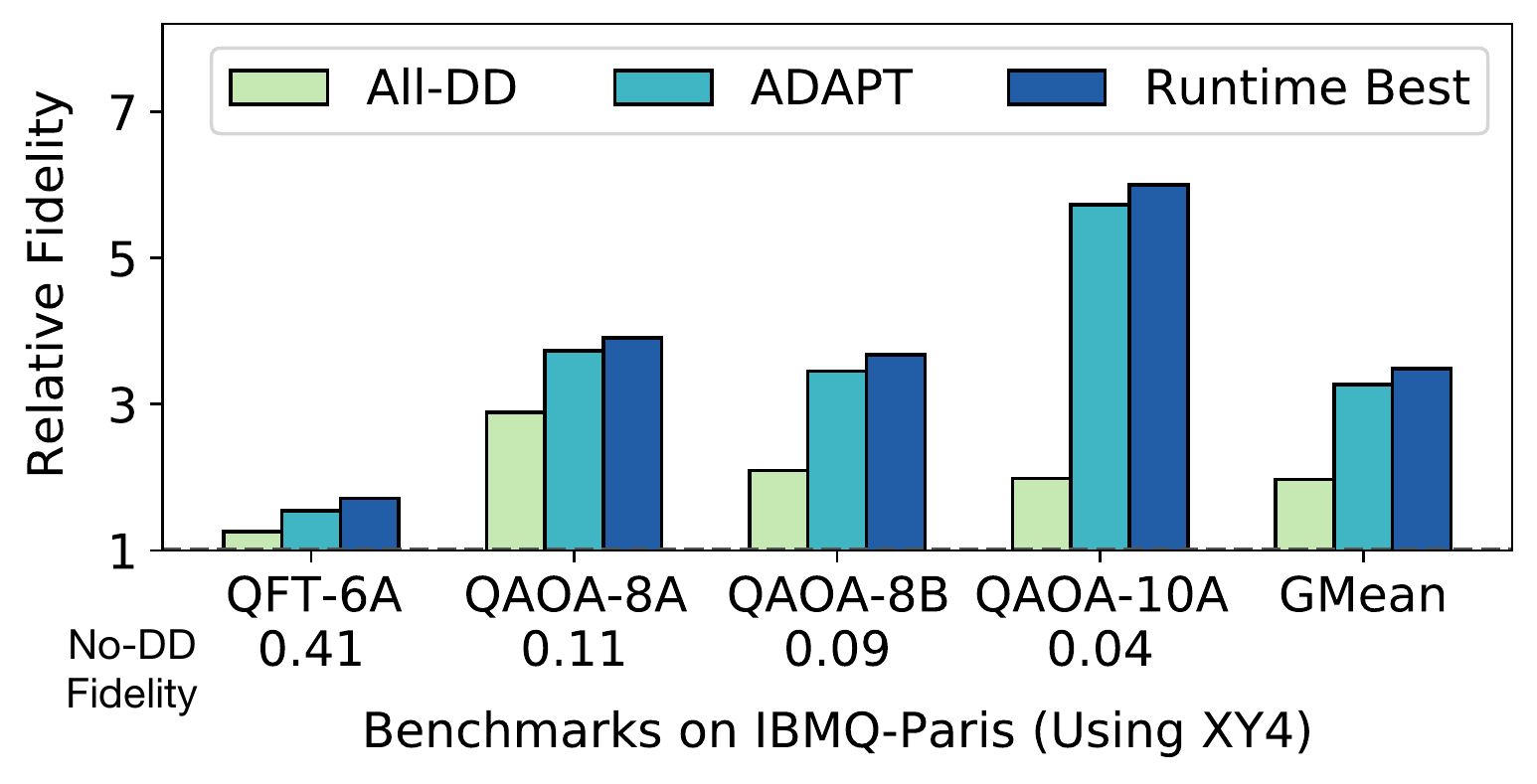}
    \caption{Relative fidelity of different policies on 27-qubit IBMQ-Paris using XY4 DD sequence.} 
    \label{fig:Paris}
\end{figure}

Figure~\ref{fig:Paris} shows the Fidelity of four benchmarks for the XY4 protocol compared to the baseline (without DD).  The number below each benchmark label specifies the baseline fidelity. We observe that on an average DD improves Fidelity. Applying DD on all qubits improves the fidelity by 1.97x on average and up to 2.89x. However, ADAPT improves the application fidelity by 3.27x and by up-to 5.7x. We also observe that the effectiveness of DD increases with increasing program size, which is expected because larger programs have more operations and depth, leaving room for longer idle time windows in the program. We also observe that the most optimal sequence at run-time outperforms both ADAPT and All-DD. This is due to the limitations of the decoy circuits and the limited search space explored by ADAPT. Our default ADAPT design searches the best sequence in a neighborhood of up-to four qubits at a time. For example, for QAOA-10, ADAPT uses only 36 decoy circuits, unlike the entire 1024 possible decoy circuits space. Nonetheless, we observe that the fidelity improvement of ADAPT is close to the runtime best and higher than All-DD for these workloads. 

We were unable to perform experiments using the IBMQ-DD protocol for IBMQ-Paris  because of changes in the basis gates and subsequent retirement of the machine.

\begin{figure*}[htp]
\centering
    \includegraphics[width=2\columnwidth]{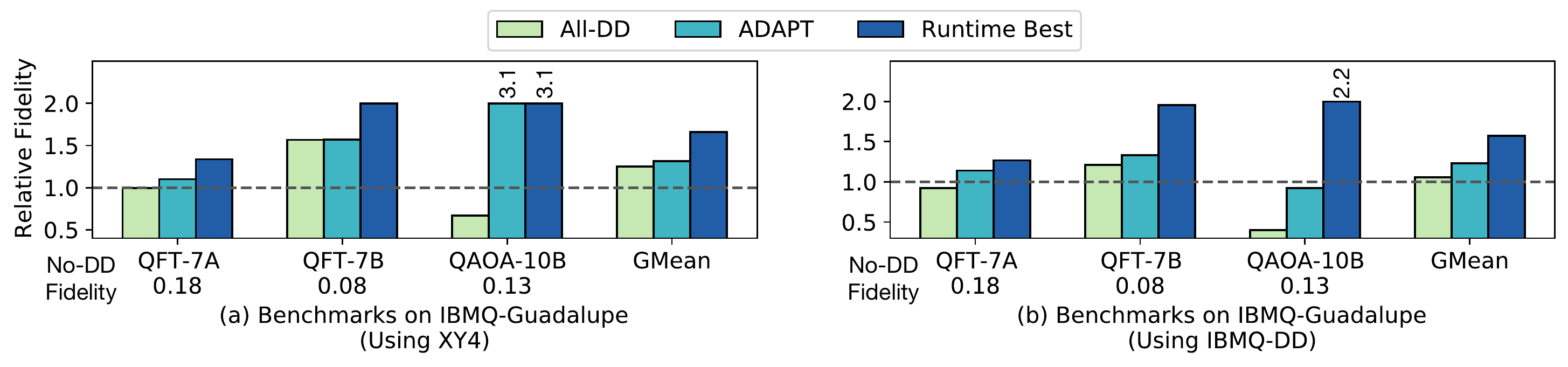}
    \caption{Relative Fidelity on 16-qubit IBMQ-Guadalupe using (a) XY4 and (b) IBMQ-DD sequences.}
    \label{fig:guadalupe_result}
\end{figure*}

\subsection{Results for IBMQ-Toronto}
Figure~\ref{fig:toronto_result} shows the Fidelity of All-DD, ADAPT and the Runtime-Best policies for 27-qubit IBMQ-Toronto machine, relative to the baseline for two different DD protocols. The number below each benchmark label specifies the baseline fidelity of the application. The structure of the QFT circuit cause qubits remain idle for substantial time periods. For example, in \texttt{QFT-6B}, Qubit-0 is idle for 90\% of the total time taken for the overall execution. Although a long sequence of DD gates adds a significant amount of single-qubit gate errors, it is still effective in improving the overall fidelity. Overall, ADAPT outperforms the baseline and improves the Fidelity by 1.52x and by up-to 3.1x for the XY4 protocol. Compared to All-DD, ADAPT improves the Fidelity on average by 1.3x and by up-to 1.89x. For the IBMQ-DD scheme, ADAPT improves the Fidelity by 1.47x and up-to 2.67x compared to the baseline. Thus, ADAPT is a generalized technique to identify qubits most vulnerable to idling errors at runtime and has applicability irrespective of the DD protocol.

\subsection{Results for IBMQ-Guadalupe}
Figure~\ref{fig:guadalupe_result} shows the Fidelity of the three different DD policies for the 16-qubit IBMQ-Guadalupe machine, normalized to the No-DD baseline. Note that this is one of the most recently released IBMQ systems with significantly reduced gate latencies and error-rates and improved coherence times. So, to test the robustness of ADAPT, we run slightly larger workloads (in terms of number of qubits, two-qubit operations, and circuit depth) on this machine. Here too, the number below each benchmark label specifies the baseline fidelity of the application. We observe that in general applying DD on all idle qubits for such large programs slightly degrades the fidelity in specific cases (\texttt{QFT-7A} for example). Note that the dominant source of errors in these circuits are not idling errors but gate and measurement errors. However, we observe that ADAPT is more robust and generally outperforms the All-DD policy. For example, 
the fidelity of the \texttt{QAOA-10B} benchmark improves by 3.1x compared to the baseline (without DD) and 4.65x compared to applying all-DD. However, we observe that in all these cases, the most optimal DD sequence at runtime outperforms the All-DD policy.

Table~\ref{tab:summary} summarizes the minimum, average, and maximum Fidelity for the three different dynamical policies normalized to the baseline on three IBMQ machines ranging from 16 to 27 qubits. Overall, ADAPT improves application Fidelity by 1.7x on average and up-to 5.73x compared to No-DD.

 \begin{table}[htp]
\centering
\begin{small}
\caption{Summary of Results}
\setlength{\tabcolsep}{0.05cm} 
\renewcommand{\arraystretch}{1.2}
\begin{tabular}{ | c || c | c | c || c | c | c|| c | c | c |}
\hline
    \multirow{2}{*}{Machine} &  \multicolumn{3}{c||}{All-DD/ XY4} &  \multicolumn{3}{c||}{ADAPT/ XY-4} & \multicolumn{3}{c|}{ADAPT/ IBMQ-DD} \\
    \cline{2-10}
    & Min & GMean & Max & Min & GMean & Max & Min & GMean & Max \\
    \hline \hline
    Paris &  1.25 & 1.97 & 2.89 & 1.55  & 3.27 & 5.73 & -- & -- &-- \\
    \hline
    Toronto & 0.58 & 1.17& 2.61& 0.68 & 1.23 &3.06  & 0.99 &1.42 &2.67 \\
    \hline
    Guadalupe &  0.67 & 1.10 & 1.57 & 1.1 & 1.31 & 3.10 & 0.92 & 1.23 & 1.33\\
    \hline 
        
\end{tabular}
\label{tab:summary}
\end{small}

\end{table}

 \ignore{
\subsection{Results for Approximation Ratio Gap}
We use Approximation Ratio Gap (ARG) to determine the effectiveness of ADAPT in the context of the QAOA benchmarks in particular. ARG determines the percentage difference between the approximation ratios on an ideal simulator and noisy outputs obtained on real hardware. A lower ARG denotes higher performance and closeness to the ideal output.~\cite{alam2020circuit}. Table~\ref{tab:eev_comparison} shows the ARG for some of the QAOA benchmarks evaluated in this paper for different machines. 

\begin{table}[htp]
\centering
\begin{small}
\vspace{-0.1 in}
\caption{Approximation Ratio Gap for XY4 Sequence (Lower is better)}
\setlength{\tabcolsep}{0.05cm} 
\renewcommand{\arraystretch}{1.25}
\begin{tabular}{ | c || c || c | c | c | c | }
\hline
    {Machine} & Program & No-DD & All-DD & ADAPT & Runtime Best \\
    \hline \hline

\multirow{4}{*}{IBMQ-Guadalupe} & QAOA-8-B & 14.88 & 19.97 & 19.32 & 12.64 \\
\cline{2-6} 
 & QAOA-10 & 17.41 & 19.47 & 13.36 & 13.36 \\
\cline{2-6} 
 & QAOA-8 & 14.62 & 19.86 & 18.20 & 13.53 \\
\cline{2-6} 
 & QAOA-9-B & 13.85 & 23.53 & 19.16 & 12.65 \\
\hline

\end{tabular}
\label{tab:benchmarks}
\end{small}

\end{table}

}

\subsection{Impact of DD Pulse Type}
\begin{figure}[b]
\centering
    \includegraphics[width=1\columnwidth]{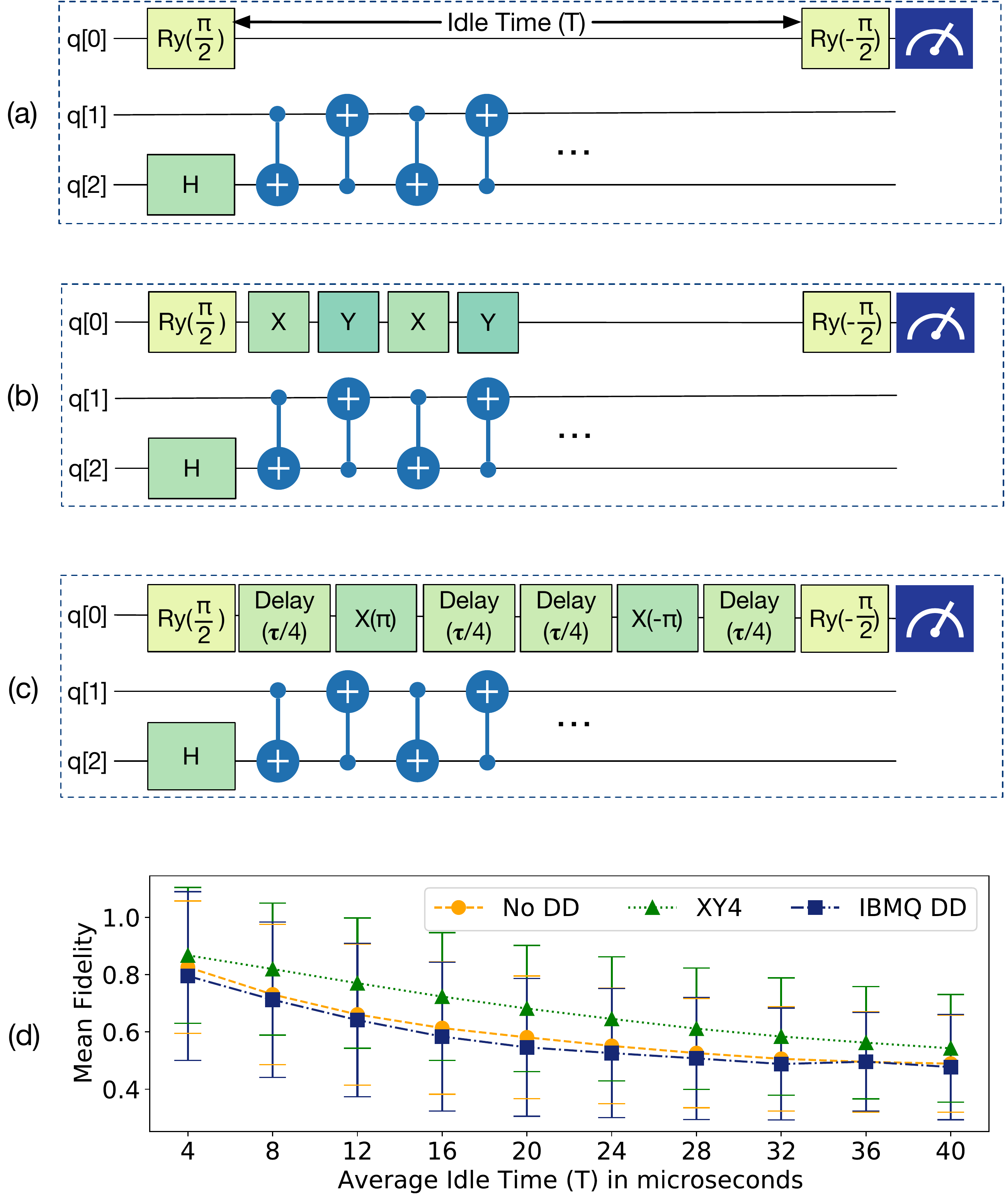}
    \caption{A characterization circuit (a) without any DD (b) with the XY4 DD sequence (c) IBMQ DD sequence. (d) Mean fidelity from individual DD sequences on 16-qubit IBMQ Guadalupe when the idle time is increased.} 
    \label{fig:dd_sequence_comparison}
\end{figure}

We compare the effectiveness of the two protocols studied in this paper standalone using some additional characterization: the XY-4 sequence and the state-of-the-art XX sequence recently proven to be effective on IBMQ systems~\cite{IBMDD}. In the experiment, we prepare three different circuits for variable idle times (T), as shown in Figure~\ref{fig:dd_sequence_comparison}. In the circuit, a quantum state is prepared by performing a single qubit rotation and the associated qubit is kept idle. Throughout the idle time, CNOT operations are repeatedly performed on a specific physical link of the device that is not connected to the qubit under study. Finally, the qubit under study is brought back to the ground state by performing the inverse rotation. In the first circuit, no DD pulses are inserted, whereas in the second circuit the XY4 DD pulses inserted throughout the idle period. The third circuit uses IBM's DD sequence in which X($\pi$) and X($-\pi$) are evenly placed by waiting for specific delay slots, as shown in Figure~\ref{fig:dd_sequence_comparison}(c). The time period for each individual delay slot is computed from the difference of idle time and length of the two X rotations, as described in Equation~\eqref{eq:delayslot}. Note that we use the most optimal decomposition for both DD pulses to ensure a fair comparison.

\vspace{0.05in}

\begin{equation}
    \label{eq:delayslot}
    \textrm{Delay }(\frac{\tau}{4}) = \frac{\textrm{T-length of X(}\pi\textrm{)- length of X(-}\pi\textrm{)}}{\textrm{4}}
\end{equation}

\vspace{0.05in}

We run these circuits for each qubit and physical link combination (total of 224 combinations possible) on 16-qubit IBMQ-Guadalupe and Figure~\ref{fig:dd_sequence_comparison}(d) shows the average fidelity of each circuit as the idle time is increased. We observe that on an average XY4 sequence outperforms the IBMQ DD sequence with increasing idle time. Similar results are reported for the Google Sycamore hardware~\cite{chen2021exponential,ai2021exponential} as well as other works~\cite{ahmed2013robustness}. This is because when idle periods are longer, there is still sufficient delay between the two X rotations during which errors can accumulate for the IBMQ-DD sequence. While such long idle periods may not be observed for random circuits used in Quantum Volume experiments, they exist in many other practical quantum applications (QFT for example). For circuits with long idling periods, we observe that the IBMQ DD protocol often performs worse than applying XY4 continuously because the latter does not experience large delays between DD pulses. To account for this in our evaluations at the application-level, we use the IBMQ-DD sequence in a more conservative manner by inserting the DD gate sequence multiple times for large idle periods. This enables a fair assessment of the DD protocol.  

\ignore{

We also study the impact of both the DD sequences at the application level when DD sequences are inserted on all the qubits and when applied to selected qubits only (optimal chosen from the runtime results obtained from all possible qubit combinations). Figure~\ref{fig:geomean} shows the geometric mean fidelity of 50 different benchmarks (up-to size 7) executed on multiple IBMQ hardware of quantum volume 32, relative to when no DD is applied. We observe that on an average XY4 performs equally well as IBM's state-of-the-art DD sequence and the optimal XY4 sequence outperforms the optimal IBMQ DD sequence \textcolor{red}{placeholder image for now}.
\begin{figure}[htp]
\centering
    \includegraphics[width=1\columnwidth]{micro54-latex-template/Figures/all_machines_dd_geomean_relative_data.pdf}
    \caption{Average impact of XY4 and IBMQ XX DD sequences on application fidelity when DD is applied to all qubits vs. when selectively applied to the best possible qubit combination (based on runtime data)} 
    \label{fig:geomean}
\end{figure}

In general, DD and ADAPT can be implemented with other pulses as well. Many advanced DD sequences exist in addition to the two sequences studies above, but most require custom pulse control or longer pulse duration, which limits applicability. For example, the XY-4 pulse latency is 320 nanoseconds, so for any idle slot larger than 320 nanoseconds, ADAPT can insert XY-4 to mitigate idling error. In theory, and for single qubit experiments, concatenated DD (CDD) pulses like XY-16 (XYXY-YXYX-XYXY-YXYX) are more effective in decoupling phase noise~\cite{souza2011robust}. However, XY-16 requires a latency of 1.6 microseconds, which means these pulses can only be applied for idle periods that are longer than 1.6 microseconds. In most NISQ applications, significant number of idle slots tend to have time duration less than one microsecond. Thus, DD protocols with long pulse duration are not as appealing for NISQ applications.  }
\newpage
\section{Related Work}

\noindent{\textbf{NISQ Compilers.}} With the increasing number of qubits and improving device quality on near-term quantum architectures, quantum software can play a vital role in improving the reliability of NISQ applications~\cite{CCS,FCNature,NAE}. To improve the application fidelity, existing quantum compilers focus on minimizing the number of gates~\cite{CGO,guerreschi,li2018tackling,DACWille} by searching for the best possible qubit mappings and sequence of SWAP operations. Moreover, recent works on noise-aware compilation use the underlying error characteristics of the hardware to avoid specific physical qubits and links such that the impact of worst-case errors on the application fidelity is reduced~\cite{tannu2018case,murali2019noise,finigan2018qubit,nishio,murali2019full,patel2020disq,patel2020ureqa,patel2021qraft,murali2019noiseadaptive,micro1,micro3}. Other works specifically target reducing certain types of errors such as crosstalk errors between ongoing CNOT operations, measurement errors~\cite{FNM,murali2020software,li2018tackling}. However, to the best of our knowledge, no compiler optimization so far focuses only on mitigating idling errors at the application level. As programs grow in size, the vulnerability of programs to idling errors increase and our paper focuses on tackling these errors.

\vspace{0.1in}
\noindent{\textbf{Dynamical Decoupling:}}  This is a well-studied noise mitigation technique that applies to a variety of qubit technologies~\cite{DDLidarAPS,DDLidarPeriodic,oliver,PhysRevLett.82.2417,paz,souza2011robust}. DD is ubiquitously used in qubit benchmarking and other small-scale experiments. More recently, Tripathi et al. have discussed the presence of crosstalk due to inter-qubit couplings and the role of DD in suppressing these errors on superconducting qubits~\cite{tripathi2021suppression}. However, this study does not consider the crosstalk from CNOT operations which exists at the application level. Also, a generic theoretical framework for DD exists that can be used to integrate DD with fault-tolerant and noise mitigation protocols~\cite{paz}. While effective for calibration and well-studied from a theoretical perspective, the trade-offs in using DD at the application-level are not fully studied.

\vspace{0.1in}
\noindent{\textbf{Dynamical Decoupling for NISQ:}}
Our work is inspired by the experimental demonstration of DD on IBM and Rigetti hardware in suppressing phase errors and improving coherence time (T2) using XY-4 and XX pulses~\cite{pokharel2018demonstration,IBMDD}. However, the use of DD to mitigate idling errors due to operational crosstalk, especially crosstalk generated by long latency CNOT gates, is not well explored. The recent study on IBMQ hardware only considers crosstalk from inter-qubit ZZ couplings~\cite{tripathi2021suppression}.  

Recently, IBM demonstrated a milestone achievement of reaching a quantum volume of 64~\cite{IBMDD} and subsequently 128 for their superconducting machine, with DD as one of the components used in benchmarking the system. IBM used a custom DD protocol, wherein the compiler inserts DD pulses on all qubits wherever possible. This is similar to the {\em DD-on-All Qubits} configuration that we study as one of our baselines.  We show that ADAPT is more effective than DD-on-All. DD on all qubits is also used in the study of quantum error correction codes (more specifically repetition codes) experiments on Google Sycamore device~\cite{ai2021exponential}.

Strikis et al.~\cite{arm2020learningbased} proposed an error mitigation strategy that inserts an extra gate before and after each operation to reduce both active and idle errors. They propose a learning scheme to identify the type of extra gates for error mitigation.  Similarly, Zlokapa et al.~\cite{alex2020deep} propose to train a deep neural network to learn the noise characteristics of a 5-qubit machine and use this network to identify the best DD pulses. ADAPT is orthogonal to these existing approaches in that it tries to identify the subset of qubits that should use DD at runtime depending on the program and device characteristics. ADAPT can use the best DD sequence for that subset of qubits. While we use decoy circuits to find the best DD sequence in ADAPT, decoy circuits have been used in other scenarios as well, such as verifying cryptographic protocols~\cite{DecoyPR}.


\section{Conclusion}
The quality and size of near-term quantum computers is improving. However, the device error-rates are still quite high and limit the fidelity of applications executed on them. In addition to facing errors while performing gate or measurement operations, qubits can also accumulate errors while remaining idle. Such idling errors present significant challenges in executing large programs. 
In this paper, we focus on mitigating idling errors for NISQ applications.

Prior works have used {\em Dynamical Decoupling (DD)} to reduce idling errors by applying a sequence of DD gates when the qubit is idle. While DD has been shown to be effective at a small scale, its applicability at the application level is not yet fully studied. We show that applying DD to all the qubits in a program is sub-optimal and may even degrade the application fidelity in some specific cases because DD is implemented by introducing additional quantum gates. If the collective error-rate of these additional operations surpasses the idling error-rate, DD can adversely impact the overall fidelity at the application-level. Thus, to reduce the impact of idling errors at the application level, dynamical decoupling must be applied judiciously. We propose a software framework {\em Adaptive Dynamical Decoupling (ADAPT)} to identify the subset of qubits that provides the highest reliability with DD. ADAPT uses decoy circuits and a localized search algorithm to perform a trial-and-error search to identify the best subset of qubits to apply DD.  We evaluate ADAPT on three quantum computers from IBM using two types of DD protocols and show that ADAPT improves fidelity by 1.86x compared to not using DD on average and by up-to 5.73x. Compared to DD on all qubits, ADAPT improves the fidelity of applications by 1.2x.

\begin{acks}
We thank Cody Jones for helpful discussions and feedback. Poulami Das was funded by the Microsoft Research PhD Fellowship. We also thank Sanjay Kariyappa and Keval Kamdar for editorial suggestions. This research used resources of the Oak Ridge Leadership Computing Facility at the Oak Ridge National Laboratory, which is supported by the Office of Science of the U.S. Department of Energy under Contract No. DE-AC05-00OR22725.
\end{acks}

\bibliographystyle{ACM-Reference-Format}
\bibliography{sample-base}


\end{document}